\documentclass[aps,prx,twocolumn,superscriptaddress,10pt,footinbib]{revtex4-2}
\usepackage{epsfig,amssymb,amsmath,mathrsfs,amsfonts,color,amsthm,graphicx,float,color}
\usepackage[active]{srcltx}
\usepackage{subfigure}
 \def\map#1{\mathcal #1}
\def\<{\langle}\def\>{\rangle}

\def\Tr{\operatorname{Tr}}\def\:{\hbox{\bf
    :}}

\def\R{\mathbb R}
\def\N{\mathbb N}
\def\C{\mathbb C}

\def\Var{\mathsf{Var}}

\def\spc#1{\mathcal{#1}}

\def\set#1{\mathsf{#1}}

\def\Proof{{\bf Proof.}}  

\newcommand*{\YY}[1]{{\color{cyan} [YY: #1]}}

\newtheorem{theo}{{Theorem}}
\newtheorem{lemma}{{Lemma}}
\newtheorem{prop}{{Proposition}}
\newtheorem{cor}{{Corollary}}

\begin{document}
\title{
 Fundamental energy requirement of reversible quantum operations}

\author{Giulio Chiribella}
\affiliation{Department of Computer Science, The University of Hong Kong, Pokfulam Road, Hong Kong}
\affiliation{Department of Computer Science, University of Oxford, Parks Road, Oxford, UK}
\author{Yuxiang Yang}
\affiliation{Institute for Theoretical Physics, ETH Z\"urich, 8093 Z\"urich, Switzerland} 
\author{Renato Renner}
\affiliation{Institute for Theoretical Physics, ETH Z\"urich, 8093 Z\"urich, Switzerland} 

\begin{abstract} 
 Landauer's principle asserts that any computation has an unavoidable energy cost that  grows proportionally to its degree of logical irreversibility. 
 But even a logically reversible operation, when run on a physical processor  that operates on different energy levels, requires energy. Here we quantify this energy requirement, providing upper and lower bounds that coincide up to a constant factor. We derive these bounds from a general quantum resource-theoretic argument, which implies that the initial resource requirement for implementing a unitary operation within an error~$\epsilon$ grows like $1/\sqrt \epsilon$ times the amount of resource generated by the operation. Applying these results to quantum circuits, we find that their energy requirement can, by an appropriate design, be made independent of their time complexity. 
\end{abstract}
\maketitle

\medskip



\noindent\emph{Introduction.} Landauer's tenet ``Information is physical"~\cite{landauer1961irreversibility} is a powerful reminder that all information processing systems are necessarily subject to the laws of physics. These laws impose certain fundamental limitations. For example, the laws of quantum theory imply  that perfect universally programmable quantum processors cannot exist~\cite{nielsen2000quantum}. Refinements of no-go results like this showed that they can be phrased as tradeoffs between the accuracy with which the tasks can be carried out and the amount of resources available for their implementation. For example, the refinement of the above-mentioned no-programming theorem asserts that the size of an approximate universally programmable quantum processor grows proportionally to $\log(1/\epsilon)$ where $\epsilon$ quantifies the tolerated error~\cite{perez2006optimality,kubicki2019resource,yang2020optimal}.

Here we consider the fundamental energy requirement for implementing quantum operations. Such requirement consists of at least two different contributions, which are consequences of the second law of thermodynamics and of energy conservation in quantum mechanics, respectively. The fact that the second law of thermodynamics has implications for the energy cost of computation is known as Landauer's principle~\cite{landauer1961irreversibility}. It asserts that any physical device that carries out a logically irreversible operation dissipates a certain minimum amount of  energy as heat, and that this amount is proportional to the degree of irreversibility (which may be quantified in terms of entropic quantities, see \cite{sagawa2009minimal,reeb2014improved,faist2015minimal,faist2018fundamental}).

In this work we are concerned with the second fundamental contribution to the energy bill.
This contribution can be regarded as a consequence of energy conservation, when applied to coherent transitions across states of different energy.
If a process is executed on a system with non-degenerate energy levels then energy must be temporarily borrowed from a \emph{battery}. For general quantum processes, this borrowing may occur in a superposition, i.e., the system's quantum state may consist of one branch in which energy has flown from the battery and another one in which energy has flown into the battery. To ensure that this does not lead to decoherence, the corresponding energy states of the battery must be indistinguishable. This, in turn, is only possible if the battery is large enough. Determining the corresponding energy requirement is exactly the topic of this work.

Previous approaches to quantify the energy requirement are based on the Wigner-Araki-Yanase (WAY) theorem~\cite{wigner1952messung,araki1960measurement}, which states that any conservation law limits the accuracy with which quantities that do not commute with the conserved quantity can be measured.
The theorem implies a bound on the \emph{variance} of the energy in the initial state of the battery required to implement an operation~\cite{ozawa2002conservative,ozawa2003uncertainty,gea2006minimum,karasawa2007conservation,karasawa2009gate,tajima2018uncertainty}. 
However, the variance does not in general provide a good bound on the size of the battery, nor on the average energy that needs to be initially stored in it.  As a simple example, consider a system with $d$ equally spaced energy levels  $\{E_0,  \dots, E_{d-1}\}$. A pure state in the superposition of energy eigenstates corresponding to $E_0$ and $E_{d-1}$ with amplitudes $\smash{\sqrt {1-1/d}}$ and $\smash{\sqrt{ 1/d}}$, respectively, has average energy (measured relative to $E_0=0$) less than $E_1$ and large energy variance that grows as $d$. On the other hand, even if the variance is fixed the energy can still take an arbitrarily large value. 

\begin{figure}[t!]
\centering
 \includegraphics[width=0.9\linewidth]{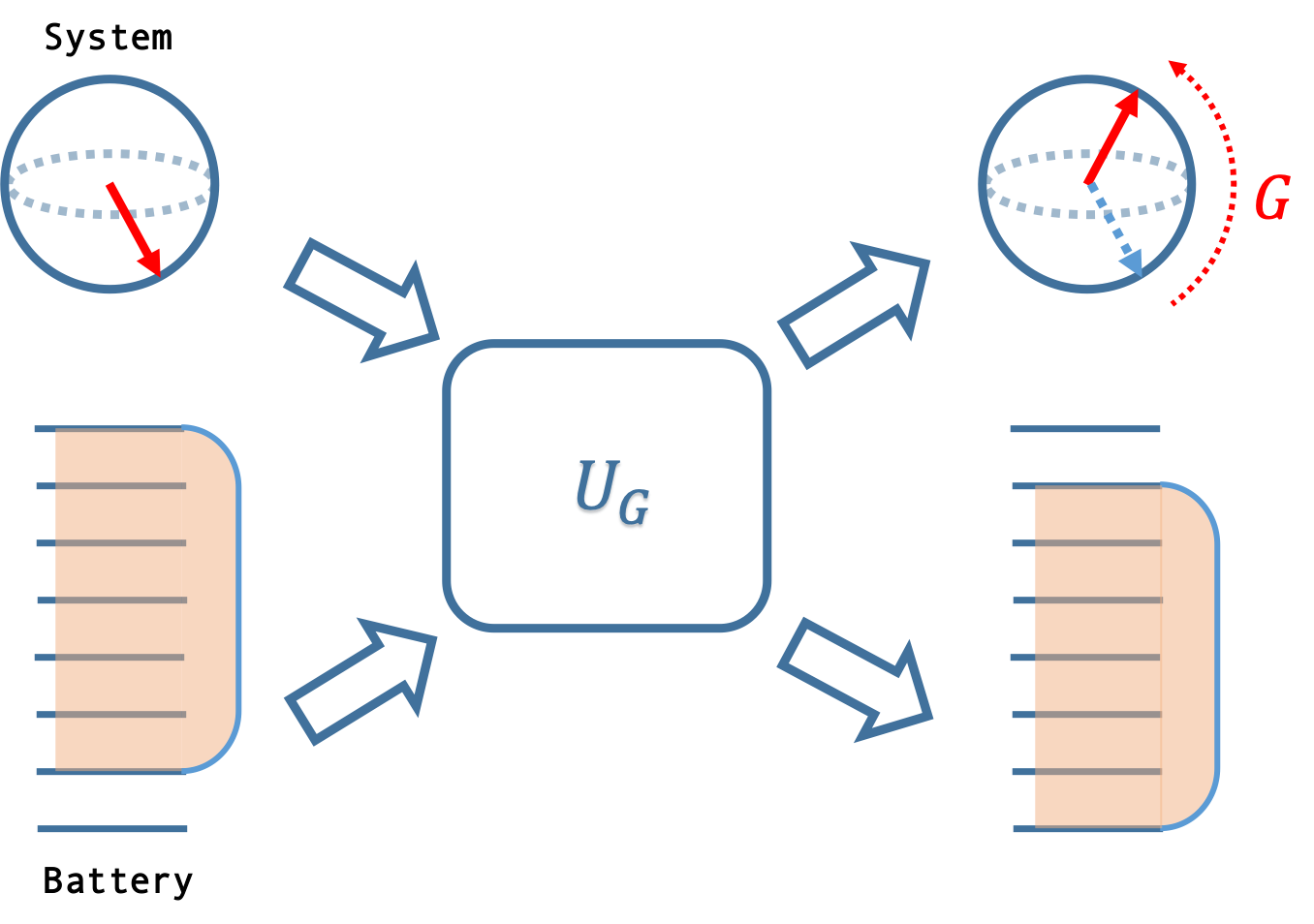}
\caption{  {\bf Implementing a reversible gate using a battery. } This figure describes a scheme that approximates a generic quantum gate $G$ that may not preserve a generic type of resource, e.g., energy, on a system. The scheme works by using a battery system that undergoes a free unitary $U_G$ together with the system. With the resource supplied by the battery, $G$ is approximated on the system.}
\label{fig:scheme}
\end{figure}

Here we instead take a general resource-theoretic approach.  Let $M$ be a function that quantifies the value of the different possible states of a system with respect to a resource. For example, $M(\rho)$ may be the average energy of the system when it is in state $\rho$. Furthermore, for a reversible operation $G$ on the system, we denote by $M(G)$ the maximum increase of the function $M$ when evaluated on an input state and on the corresponding output state produced by $G$. Hence, in the case where the considered resource is energy,  $M(G)$ quantifies by how much the system's energy can grow when executing $G$. Assume now that we want to implement $G$ up to a precision $\epsilon$ (which we quantify in terms of the worst-case infidelity, defined below). The implementation should consist of a device that can merely carry out \emph{free} operations, i.e., operations that cannot generate the resource. Such an implementation must necessarily use a battery, as illustrated in Fig.~\ref{fig:scheme}. Then the following general assertion can be made. 

\begin{theo}\label{thm-resource}
If the resource measure $M$ is monotonous, additive, and regular (see later for definitions) then every approximation of a reversible operation $G$ within error~$\epsilon$ using a free device $U_G$ connected to a battery in state  $\beta$ must satisfy  
\begin{align*}
M(\beta)\ge\frac{\left(M({G})+M({G}^{\dag})\right)^2}{32K_{\rm  S}\sqrt{\epsilon}} - c-O(\sqrt{\epsilon})\, ,
\end{align*}
where $c$ and $K_{\rm S}$ are constants that merely depend on $M$ and the system on which $G$ acts. 
\end{theo}



This theorem, whose proof we will sketch in the first section below, yields in particular  a lower bound on the energy requirement for implementing a reversible operation $G$. Specifically, in the second section, we show that the average energy content $\<H_{\rm B}  \>$ of the battery supplying energy to the processor must be at least
\begin{align}\label{energybounds}       \<  H_{\rm B}  \>   \ge  \frac{\left((\lambda_{\max}-\lambda_{\min})(\Delta_G H_{\rm S})\right)^2 }{  32 \sqrt \epsilon  \,  \| H_{\rm S}\|}  -O(\sqrt{\epsilon})\, , 
\end{align}
where $\| H_{\rm S}\|$ is the operator norm of the system's Hamiltonian, $\lambda_{\max}$ ($\lambda_{\min}$) denotes the maximal (minimal) eigenvalue, and $\Delta_G H_{\rm S}=    G^\dag H_{\rm S}   G  - H_{\rm S}$ is the change of the system's Hamiltonian  induced by the action of the gate $G$.
We have assumed, without loss of generality, that the minimum energy is zero for both the system and the battery, and thus $\|H_{\rm S}\|$ is equal to the maximum energy of the system.
The bound  (\ref{energybounds}) states that the average energy of the battery should be above the ground state energy  by an amount determined by the energy change operator $\Delta_G H_{\rm S}$, the system's energy scale  $\| H_{\rm S}\|$,  and the error $\epsilon$.  

While the bound~(\ref{energybounds}) depends on the particular operation $G$, by maximising over all such reversible operations we obtain a bound on the energy requirement of a universal quantum processor operating on a system $S\rm $ with a given Hamiltonian $H_{\rm S}$, 
\begin{align}\label{lower}
\<  H_{\rm B}  \>    \ge     \frac{\| H_{\rm S}\| }{  8 \sqrt \epsilon    }   - O(\sqrt{\epsilon}) \, . 
\end{align}
This bound is tight up to a constant factor. More precisely, assuming that the system has equally spaced energy levels, we show by an explicit construction, described in the third section below, that 
\begin{align}\label{upper}
\<  H_{\rm B}  \>  & \le  \frac{   \pi  \,   \| H_{\rm S}\|}{ 2\sqrt{ \epsilon}}  \,.  
  \end{align}  
Taking together these two bounds, we have thus established that the fundamental energy requirement for operating on $\rm S$  grows as $ \| H_{\rm S}\|/\sqrt \epsilon$. 
Note that if the system's Hamiltonian is fully degenerate, i.e., $\| H_{\rm S}\|=0$ then energy conversation does not  imply an energy requirement. Besides the average energy, we show that the energy spread of the battery is lower bounded by $ \| H_{\rm S}\|/\sqrt \epsilon$, and the tightness of the bound can again be achieved with the construction that led to~(\ref{upper}). 

\begin{figure}[t!]
\centering
\subfigure[]{\label{fig:multiple}
 \includegraphics[width=0.8\linewidth]{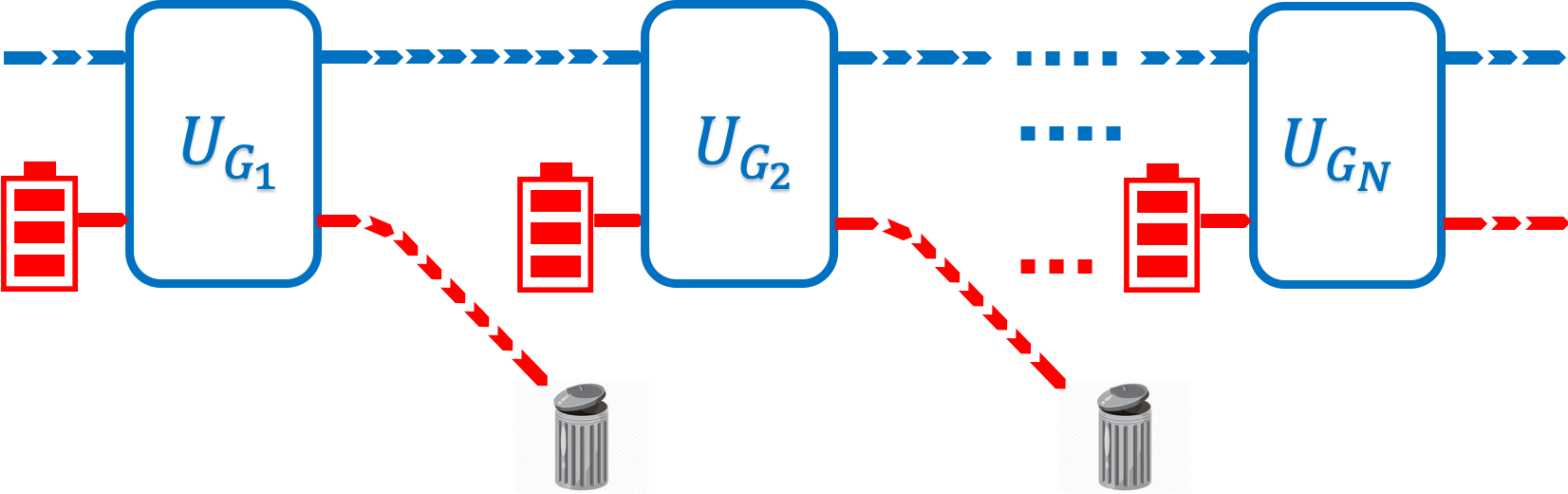}
}
\hfill
\subfigure[]{\label{fig:single}
\includegraphics[width=0.8\linewidth]{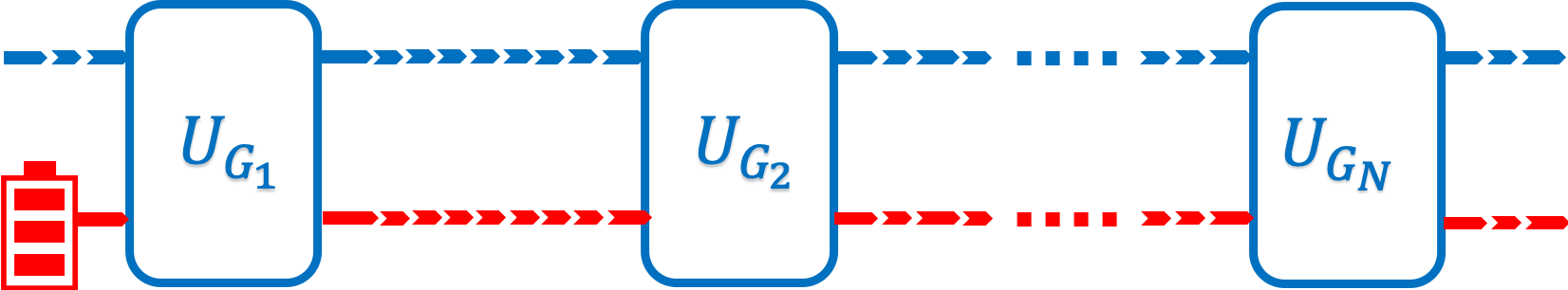}
}

\smallskip

\caption{  {\bf Multiple and single battery implementations of quantum computation.} Two different setups of energy-preserving quantum computation are compared. Fig.\ \ref{fig:multiple} depicts the multiple battery implementation that has often been considered in previous work \cite{ozawa2002conservative,ozawa2003uncertainty,gea2006minimum,karasawa2007conservation,karasawa2009gate,ikonen2017energy}, where each single gate of the circuit is equipped with an individual battery that is discarded after the gate is implemented. In contrast, in this work we consider the single battery implementation as illustrated in Fig.\ \ref{fig:single}, where a single battery provides energy for the whole circuit and is reused after the implementation of each individual gate.
}
\label{fig:circuit}
\end{figure}

Finally, we determine how the energy requirement of a quantum circuit depends on its complexity. Previous works considered  implementations of quantum circuits where each gate is powered by an independent battery \cite{ozawa2002conservative,gea2006minimum} (see Fig.~\ref{fig:multiple}).    The energy requirement  then obviously grows linearly with the number of non-conservative gates, making complex computations energetically demanding.   In contrast, we show that the energy requirement of  quantum circuits is independent of their complexity.  
For this we consider an implementation that uses a single battery to power all gates in the circuit (see Fig.~\ref{fig:single}). It turns out that energy can be recycled from one gate to the next, and that the energy requirement for a sequence of gates is exactly equal to the energy requirement of the overall gate resulting from their composition.  Hence, the energy requirement depends only on the size of the computational register, but not on the time complexity of the computation.  For quantum computations with classical inputs and outputs, such as Shor's algorithm, we further show that our implementation  is exact and the energy requirement is just the energy needed to write down the output of the computation. This may be regarded as the quantum analogue of a classical result by Fredkin and Toffoli \cite{fredkin1982conservative}, who studied the fundamental energy constraints that the classical laws of physics impose on computation.

\medskip
\noindent\emph{Lower bound on general resource requirement.} 
 In this section we spell out the assumptions underlying Theorem~\ref{thm-resource} and describe the main proof idea. (The full proof is provided in Appendix~\ref{app-mainthm}.) For this we take a resource-theoretic viewpoint, i.e., 
we start from a given set of  \emph{free} operations that is closed under composition~\cite{coecke2016mathematical}.  Let $U_G$ be such a free operation that acts on both a system $\rm S$ and a battery $\rm B$, which is initialised in state $\beta$. The resulting operation on $\rm S$ is then described by the quantum channnel
\begin{align}\label{approx}
\map E_G (\cdot)   =    \Tr_{\rm B}  \left[   U_G  (\cdot  \otimes  \beta )   U_G^\dag\right] \, ,
\end{align} 
where $\beta$ is the initial state of the battery and $\Tr_{\rm B}$ denotes the partial trace over the battery's Hilbert space.


To quantify how well the operation $\map E_G$  approximates a desired gate $G$ we use the \emph{worst-case fidelity} $F_{\rm wc}$ between the output of  $G$  and~$\map E_G$ for any input, which may also be correlated to an external reference system ${\rm R}$. That is, explicitly,
\begin{align}\label{wc-fidelity}
F_{\rm wc}:=\inf_{\rm R}\inf_{|\Psi\>  \in  \spc H_{\rm S} \otimes \spc H_{\rm R}} \Tr\left[(\map E_{G} \otimes \map I_{\rm R})  (  \Psi ) \,   (\map{G} \otimes \map{I}_{\rm R})(\Psi)   \right] \, ,
\end{align}
with $\Psi : = |\Psi\>\<\Psi|$, $\map G  (\cdot)  =  G \cdot G^\dag$, and  $\map I_{\rm R}$ denoting the identity map on $L(\spc H_{\rm R})$, the space of linear operators on $\spc H_{\rm R}$. We say that an implementation has  error $\epsilon$ if $F_{\rm wc}=1-\epsilon$.  The use of this error measure is justified by the fact that the resource requirements,  in the case of energy as discussed in the introduction, can be bounded tightly in terms of~$\epsilon$ (up to a constant). We also note that the fidelity is easy to evaluate and   widely used to quantify the quality of gates in quantum computation. Moreover, it may be related to other measures of distance, e.g., the diamond norm \cite{kitaev1997quantum} (see~\cite{diamondnote} for a definition) via the inequalities $1-\sqrt{F_{\rm wc}}\le \frac12\|\map{E}_{G}-\map{G}\|_{\diamond}\le\sqrt{1-F_{\rm wc}}$. In Appendix~\ref{app-diamond}, we show that the dependence of the energy requirement on the diamond norm also scales as $1/\sqrt{\epsilon}$, up to a factor that may however depend on the system's dimension.

Theorem~\ref{thm-resource} is a general resource-theoretic statement, which merely depends on general properties of the measure $M$
 used to quantify resourcefulness.
Specifically, for any given system, $M$ is a function of the density operator of that system such that the following holds:
  \begin{enumerate}
  	\item {\em Monotonicity.} $M$ is non-increasing under free operations and partial trace.
  	\item {\em Additivity on product states.} $M  (\rho \otimes \sigma) =  M  (\rho)  +  M(\sigma)$. 
  	\item {\em Regularity.}  There exists a constant $c\in \R$ and, for any system  $\rm S$, a Lipschitz constant  $K_{\rm S}\ge 0$,  such that $|M(\rho)-M(\sigma)|\le  K_{\rm S} \, \|\rho-\sigma\|_1+c$ for any states $\rho$ and $\sigma$ of system $\rm S$, and such that $K_{\rm S}$ is subadditive, i.e., $K_{\rm AB} \le K_{\rm A}+K_{\rm B}$ for any systems $\rm A$ and $\rm B$.
	\end{enumerate}

With these definitions in place, we can now proceed to the proof of  Theorem~\ref{thm-resource}.  Let  $\map V_G  (\cdot)   :=   \map U_G (\cdot \otimes  \beta  )$ with  $\map U_G  (\cdot )  =  U_G \cdot U_G^\dag$ be the evolution defined in~\eqref{approx}, but before tracing out the battery~$\rm B$. Using techniques from \cite{kretschmann2008information,chiribella2013short,gutoski2017fidelity} we show (see Appendix~\ref{app-mainthm}) that the channel $\map V_G$ is close to $\map G \otimes \beta'$, where $\beta'$ is a suitable battery state \cite{2footnote}.  Due to its additivity property, it is useful to measure this closeness in terms of the diamond norm $\| \cdot \|_\diamond$~\cite{diamondnote}
\begin{align}\label{stine1}
 \|  \map V_G       -  \map G\otimes \beta'  \|_\diamond  \le 2\sqrt \epsilon \,.      
 \end{align} 
But this means that approximately there is no entanglement between the system and the battery after the evolution, and the battery ends up in a state close to  $\beta'$. Conversely, the state  $\beta'$ may be used to approximately implement the inverse gate $G^\dag$, using the gate $U_{G}^\dag$, i.e., 
\begin{align}\label{stine2}
 \|   \map G^{-1}\otimes \beta-   \map V'_G  \|_\diamond  \le 2\sqrt \epsilon \, ,         
\end{align}  
with $\map V'_{G}  (\cdot)   :=    \map  U_G^{-1}   (  \cdot  \otimes \beta')$. 

\begin{figure}[t!]
\centering
 \includegraphics[width=\linewidth]{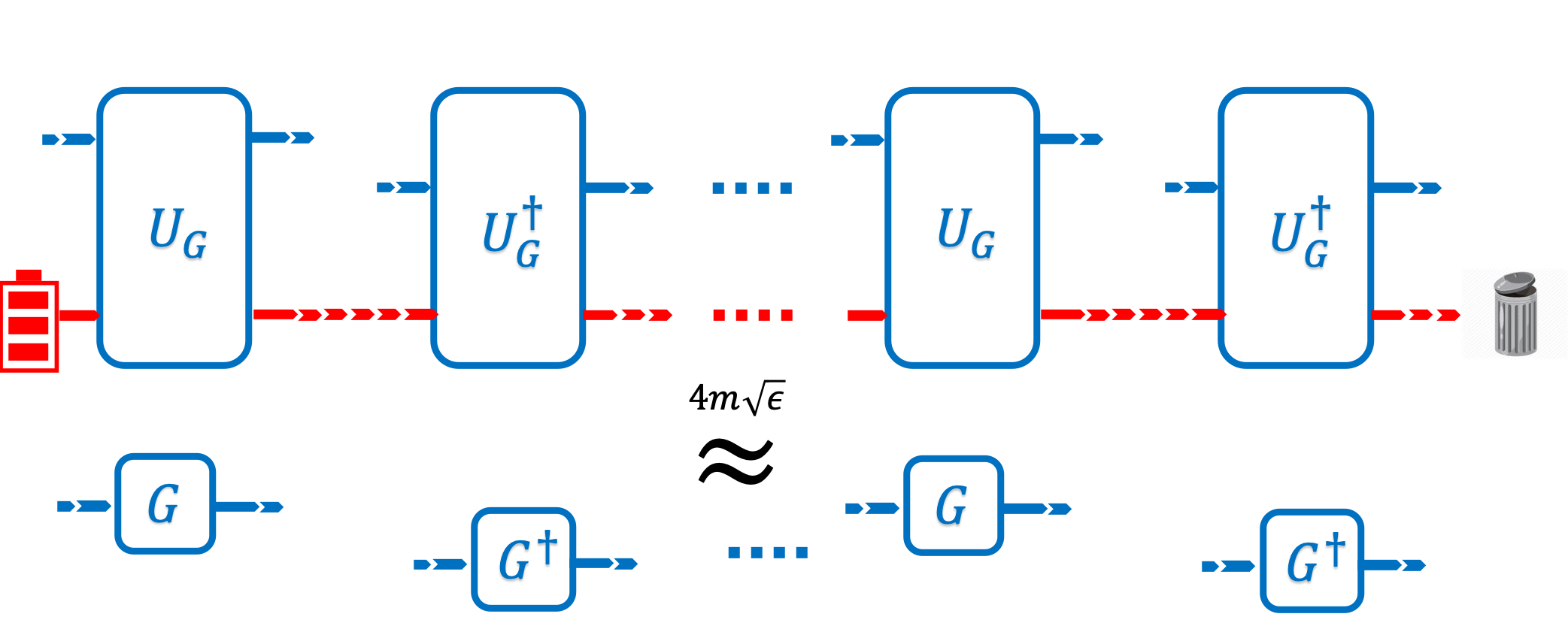}
\caption{  {\bf Approximating $m$ uses of a gate and its inverse. } If a unitary gate $G$ can be implemented with small error, then the battery can be reused  $2m$ times, approximately implementing $m$ uses of the gates  $G$ and $G^\dag$. As a consequence, the state of the battery should be able to provide $m$ times the maximum resource generation of $G$ and $G^\dag$ up to a correction. 
}
\label{fig:comb}
\end{figure}
\
According to~(\ref{stine1}) and~(\ref{stine2}) we may thus implement the gate $G^\dag$ after gate $G$, thereby returning the ancilla approximately to its initial state. Repeating this procedure $m$ (for any $m \in \mathbb{N}$) times, i.e., 
composing $m$ implementations of $G$ and of $G^\dag$ in alternating order as illustrated in Fig.~\ref{fig:comb}, we still approximate each of them within  an error bounded by $4 m \sqrt {\epsilon}$. Note that the circuit in the lower half of the figure increases the resource value $M$ by virtue of $G$ and $G^\dag$. To approximate this increase of resource, the circuit in the upper half of the picture must use the battery, because all the other operations in the circuit are free and therefore resource non-generating. 
Hence, the amount of resource generated by $m$ uses of $G$ and $G^\dag$, i.e., $m$ times $M( G) + M(G^{\dag})$, must be matched by the battery. Evaluating this amount leads to the bound stated in Theorem~\ref{thm-resource}.

\smallskip

\noindent\emph{Lower bound on energy requirement.} 
To obtain~(\ref{energybounds}) we apply Theorem~\ref{thm-resource} to the resource theory where the free operations are energy-preserving channels \cite{chiribella2017optimal,sparaciari2017resource}.  
For any system, the resource function $M$ is defined by $M(\rho)=\Tr[H\rho]$, where $H$ is the system's Hamiltonian 
(with the minimum energy set to zero) and $\rho$ is the system's density operator.  This resource function is additive on product states and it is non-increasing under energy preserving channels and partial trace. 
Moreover, the inequality $|\Tr(\rho-\sigma)H|\le\|\rho-\sigma\|_1\cdot\|H\|$ shows that the function $M$ is Lipschitz continuous with Lipschitz constant  $K_{\rm S}=\|H\|$, equal to the energy scale of the system under consideration, and $c=0$. Finally, we note that $M(G)=\max_{|\psi\>\in\spc{H}_{\rm S}}\<\psi|G^\dag H_{\rm S}G|\psi\>-\<\psi|H_{\rm S}|\psi\>=\lambda_{\max}(\Delta_G H_{\rm S})$ and $M(G^\dag)=\max_{|\psi\>\in\spc{H}_{\rm S}}\<\psi|G H_{\rm S}G^\dag|\psi\>-\<\psi|H_{\rm S}|\psi\>=-\lambda_{\min}(\Delta_G H_{\rm S})$, where $\lambda_{\max}$ ($\lambda_{\min}$) denotes the maximal (minimal) eigenvalue. Inserting all this into Theorem~\ref{thm-resource}, we immediately obtain the desired bound~\eqref{energybounds} on the average energy content $\<H_{\rm B}\> = \Tr[H_{\rm B} \beta]$ of the battery $\rm B$ required to implement~$G$. 

In a similar way we can also  derive a lower bound on the required total capacity $\|H_{\rm B}\|$ of the battery. For this we first apply Theorem~\ref{thm-resource} to the resource function $M'(\rho):=\Tr\rho(\|H\|\cdot I-H)$ to obtain the bound
\begin{align}\label{bound-energy-inter2}
\|H_{\rm B}\|-\<H_{\rm B}\>\ge  \frac{\left((\lambda_{\max}-\lambda_{\min})(\Delta_G H_{\rm S})\right)^2 }{  32 \sqrt \epsilon  \,  \| H_{\rm S}\|} - O(\sqrt{\epsilon}) \, . 
\end{align} 
Taking the worst case $G$, which satisfies $\lambda_{\max}( \Delta_G H_{\rm S})=-\lambda_{\min}(\Delta_G H_{\rm S}) = \| H_{\rm S}\|$, and combining this with bound~(\ref{energybounds}), we find a bound on the maximum energy (or the capacity) of the battery,
 \begin{align} \label{capacity-bound}
\|H_{\rm B}\|  \ge \frac{\| H_{\rm S}\|}{4\sqrt \epsilon}- O(\sqrt{\epsilon}) \ .
 \end{align}  
This and~\eqref{lower} are lower bounds on the energy requirement of a  universal processor, able to implement arbitrary gates on system $\rm S$ with error $\epsilon $ or less. While~\eqref{lower} quantifies the energy requirement in terms of the average energy that a battery must contain, \eqref{capacity-bound} refers to the battery's total capacity.

Theorem~\ref{thm-resource} also provides bounds on other types of resources, such as coherence \cite{baumgratz2014quantifying,winter2016operational,marvian2016quantify,streltsov2017colloquium}. A concrete example is the relative entropy of coherence  \cite{baumgratz2014quantifying} $C(\rho):=S(\rho_{\rm diag})-S(\rho)$, with $S$ denoting the von Neumann entropy  and $\rho_{\rm diag}$  the diagonal part of $\rho$ in the energy eigenbasis. Here the theorem yields the bound (see Appendix~\ref{app-coherence})
\begin{align}
C(\beta)\ge\frac{\left(C(G)+C(G^\dag)\right)^2}{32\sqrt{\epsilon}\log d_{\rm S}}-O(1).
\end{align}
on the initial coherence that the battery must provide, where $d_{\rm S}$ is the dimension of the system on which $G$ acts, and $C({G})$ is the amount of coherence generated by the gate $G$. For gates like the generalized Hadamard gate this quantity can be as large as $\log d_{\rm S}$. Therefore, the minimum amount of coherence required to operate a universal quantum processor scales like $\log d_{\rm S}/\sqrt{\epsilon}$.
 
 We have shown that the requirement for energy and coherence both follow an $1/\sqrt{\epsilon}$ scaling. The same scaling characterises also the standard deviation of the energy, as observed in previous works \cite{ozawa2002conservative,ozawa2003uncertainty,gea2006minimum,karasawa2007conservation,karasawa2009gate,tajima2018uncertainty}. 

\medskip
\noindent\emph{Attaining the bound.} We now show that the bound (\ref{lower}) can be attained with a suitable choice of  battery state and  interaction between the system and the battery.   In this part, we assume that the system has equally spaced energy levels,  which is the case, for example, if it consists of  $n$ identical individual qubits. We denote the spacing by $\hbar \omega$.  

The implementation uses a battery with equally spaced energy levels with spacing $\hbar \omega$, ranging from $0$ to $\|H_{\rm B}\|=R\|H_{\rm S}\|$, where  $R$ is an integer, assumed to be  larger than 2 for later convenience.     At the beginning, the battery is initialized in a superposition of energy eigenstates with sine-shaped amplitudes  \cite{buzek1999optimal}
 \begin{align}\label{sine}
  |\beta  \>=  \sqrt{\frac{2  }{L}}\sum_{E_{{\rm B}}=  \|H_{\rm S}\|}^{(R-1)  \|H_{\rm S}\|}  \, \sin\left(\frac{(E_{\rm B}-\|H_{\rm S}\|+\hbar \omega)\pi}{\hbar \omega  L}\right)  \,  |E_{{\rm B}}\>  \, ,
\end{align}
where  the summation runs  in steps of $\hbar\omega$, and $L  =  ({R  -  2})  \|H_{\rm S}\|/(\hbar\omega)    + 2$. Note that the lowest and highest energy levels are unoccupied. This allows the battery to both supply and absorb energy from the system.

For the interaction between the system and the battery we adopt  a   construction from Refs.\ \cite{skrzypczyk2013extracting,aaberg2014catalytic,navascues2014energy}, suitably adapted to unitary gates on finite-dimensional systems.  
Denote by $E_{\rm S,x}$ the energy of $|\psi_x\>$.
For a given value   $E$  of the total energy,  and for every $x$ satisfying the condition 
\begin{align}\label{match}
  E-  \|H_{\rm B}\| \le E_{{\rm S},x}  \le E
\end{align}
we define the eigenstates 
\begin{align}
|x,  E\>      :  =    |\psi_x\>  \otimes  |E-  E_{{\rm S},x}\>\, .
\end{align} 
Then, we  denote by $\set E_{\rm ok}$ the set of values of the total energy such that condition (\ref{match}) is satisfied for every $x  =1,\dots,  d_{\rm S}$, or equivalently, the set of values $E$ satisfying the condition $  \|H_{\rm S}\|   \le E \le  \|H_{\rm B}\|$.    
For every $E  \in  \set E_{\rm ok}$,  define the partial isometry  
\begin{align}
U_G^{(E)}  :  =  \sum_{x,y=0}^{d_{\rm S}-1} \,     \<\psi_x  | G  |\psi_y\>    ~  |x, E\>\<y,E|  \, ,
\end{align} 
which acts as the unitary gate $G$ in the eigenspace with total energy $E$.  To make the computation reversible  on the whole system $\rm S B$, we set $U_G$ to be the unitary gate 
\begin{align}\label{ug}
U_G  :  =    \sum_{E  \in  \set E_{\rm ok}}  \,    U_G^{(E)}   +   \sum_{E  \not \in \set E_{\rm ok}}  \,   P_E \, ,
\end{align}  
where $P_E$ is the projector on the subspace with total energy $E$.

In Appendix \ref{app-lowerbound} we show that the worst case fidelity of the above implementation is lower bounded as 
\begin{align}\label{Fwc}
F_{\rm wc}     \ge 1-\left(\frac{\pi (\lambda_{\max}-\lambda_{\min})(\Delta_G H_{\rm S})}{ 4  \<  H_{\rm B}  \>}\right)^2  \,      \left(1+O\left(\frac{\|H_{\rm S}\|}{\< H_{\rm B }\>}\right)\right)   \, , 
\end{align}
and therefore the energy requirement is upper bounded as 
\begin{align}
\<H_{\rm B}\>  \le   \frac{\pi   (\lambda_{\max}-\lambda_{\min})(\Delta_G H_{\rm S})}{4 \sqrt {\epsilon}}  \, \left(1+O\left(\frac{\|H_{\rm S}\|}{\< H_{\rm B }\>}\right)\right)   \, .  
\end{align}
In the worst case over all possible gates,  one has  $(\lambda_{\max}-\lambda_{\min})(\Delta_G H_{\rm S})=2\|H_{\rm S}\|$, matching the lower bound (\ref{lower}) up to a constant factor of $4\pi$.  
 
The error $\epsilon$ depends on the parameter $R$ that characterizes the battery state (\ref{sine}).  Observing that the energy of the sine state is  
$\<  H_{\rm B}  \>       =  R \|H_{\rm S}\|/2$, we obtain the dependency $ R  \approx   \pi   (\lambda_{\max}-\lambda_{\min})(\Delta_G H_{\rm S})/ (2\sqrt {\epsilon}\|H_{\rm S}\|)$.    Therefore, the battery capacity of this implementation is $\|H_{\rm B}\|=R\|H_{\rm S}\|\approx  \pi   (\lambda_{\max}-\lambda_{\min})(\Delta_G H_{\rm S})/(2\sqrt {\epsilon})$. Taking the worst-case $G$, the capacity of the battery is approximately
\begin{align}
\|H_{\rm B}\|\approx  \frac{\pi   \|H_{\rm S} \|}{\sqrt {\epsilon}},
\end{align}
matching the lower bound (\ref{capacity-bound}) up to a constant of $4\pi$. 


\medskip
\noindent\emph{Energy-efficient quantum computation.}  We established the minimum energy requirement of one single quantum operation. But what about a computation that consists of many individual steps?    One way to implement the computation is  to assign an individual battery to each gate and to replace the gate by its conservative approximation.   However,  this approach leads to a heavy energy toll.   If each gate is powered by an individual battery of energy    $ \<  H_{\rm B}\>$,  then bound  (\ref{lower}) implies that the error  cannot decrease faster than $1/\<  H_{\rm B}\>^2$.   The error (infidelity) is a lower bound on the trace distance, which in the worst case increases linearly with the number of gates.  The linear increase implies that at most  $O(\<H_{\rm B}\>^2)$ gates  can be combined together with tolerable error.  For a circuit of $N$ non-conservative gates, this means that the energy of each individual battery should grow at least as $\sqrt N$, with a total energy requirement scaling at least as $N^{3/2}$.   
In other words, the energy requirement depends on the number of non-conservative gates, just as in traditional models of dissipative computation. 

We now show that, in fact, quantum computation can be implemented with an amount of energy that is independent of the circuit depth. To do so, we propose a scheme of computation where energy is recycled from one computational step to the next.  
The computation is performed on  $n$ identical qubits, each with Hamiltonian $H_{\rm S}^{(1)}$ and energy gap $\|H_{\rm S}^{(1)}\|  = \hbar \omega$, and uses a single battery of capacity $C_{\rm B}  =   R n \|H^{(1)}_{\rm S}\|$,   where $R$  is an integer  depending on the desired level of accuracy.   For an elementary  gate  $G$ acting on a subset of  $k$ qubits, we let the battery and the  $k$ qubits interact through  the  energy-preserving gate $U_G$ in (\ref{ug}).   
 The energy subspaces on which the gate $U_G$ acts non-trivially correspond to the energy values $\set E_{\rm ok}^{(k)}   =\{  E  \,  |~  k  \,\|H_{\rm S}^{(1)}\|   \le E \le  \|H_{\rm B}\|\}$.   
 Now, consider the total energy of the $n$ qubits and the battery.  
 For every two gates $G_1$ and $G_2$, one has  the property 
 \begin{align}
 U_{G_1}     U_{G_2}  P_{\rm ok}^{(n)}  =  U_{G_1 G_2}  P_{\rm ok}^{(n)} \, ,
 \end{align}
 where $P_{\rm ok}^{(n)}$ is the projector on the eigenspaces of the total energy in $\set E_{\rm ok}^{(n)}$.  The above relation means that the local interactions  of the battery with  subsets of qubits    are enough to generate every  global  interaction between the battery and all the qubits involved in the computation.  Hence, the computation can be realized by preparing the battery in the state (\ref{sine}), with $\|H_{\rm S}\|  =  n  \|H_{\rm S}^{(1)}\|$.    For a computation consisting of $N$ gates $ (G_i)_{i=1}^N$,  the energy requirement does not depend on $N$, but only on the unitary $  G  =  G_N \cdots G_2G_1$ that describes the overall computation.    Since the gate $G$ acts on at most $n$ qubits, the energy requirement for implementing any computation with accuracy $\epsilon$ is at most $\pi  n \hbar\omega/(2 \sqrt \epsilon)$.

It is worth noting that, if the computation is only required to work on a subset of input states, the energy requirement can be lower. For example, suppose that a computation has classical input and classical output, as in Shor's algorithm and in many other quantum algorithms.  In this case, every computation can be implemented  {\em exactly} by setting the battery in the initial state with energy $n\hbar \omega$, and then using the interaction (\ref{ug}) for every gate (see Appendix \ref{app-computing}).   


    
\medskip 

\noindent {\em Conclusions.}   We derived a bound on the resources that are required to approximately implement a reversible quantum operation.  We found that, for a general class of resources, which include energy as a special case, the resource requirement grows as $1/\sqrt \epsilon$, where $\epsilon$ is the approximation error (Theorem~\ref{thm-resource}).  Furthermore, in the case where the resource is energy,  the bound is attainable within a constant factor, provided that the target system has equally spaced energy levels. A typical example for such a situation is a quantum processor acting on $n$ identical qubits.  For a computation, this minimum energy requirement is, remarkably, achievable even if the computation is carried out by a complex quantum circuit with many individual unitary gates.  In this case, we showed that the battery state can be recycled from one computational step to the next, making the energy requirement 
independent of how the computation is decomposed.    

 Our bound on the energy requirement is unrelated to the second law of thermodynamics: it follows from the conservation of energy, and it is present even if the evolution is entirely reversible. Nonetheless, our energy requirement can be compared quantitatively with  the thermodynamical work requirement associated to Landauer's principle, which is present when the evolution is irreversible.   Landauer's principle sets the work cost of erasing information from a single qubit to $K_B T$, where $K_B$ is Boltzmann's constant and $T$ is the system's temperature.  For superconducting qubits, assuming an operation temperature of the order of 1K, the Landauer's cost is of the order of $10^{-23}J$.  
  Our bounds~\eqref{lower} and~\eqref{upper}, on the other hand, introduce a new energy requirement that depends on the Hamiltonian of the qubit system and the desired implementation accuracy. For transmon superconducting qubits, the energy gap between $|0\>$ and $|1\>$ is around the order of 10 GHz \cite{koch2007charge,kjaergaard2020superconducting}, implying an energy requirement of $\sim 10^{-24}\times \epsilon^{-\frac12}J$. 
  The energy requirement is thus comparable to the energy cost predicted by Landauer's principle. 

Like Landauer's principle, our results must be understood as fundamental limitations imposed by the laws of physics. At least for today's few-qubit devices, which require large cooling and control machinery external to the actual quantum processors, the fundamental energy requirement  as given by~\eqref{lower} and~\eqref{upper} merely represents a minor part of the overall energy consumption. However, as quantum technology is being developed further, the energy required, e.g., for cooling, will most likely  scale less than linearly with the number of qubits, and its contribution to the overall energy bill thus become less dominant.  Analogously to how the fundamental bounds of classical thermodynamics have helped us optimising engines, a theory of the \emph{thermodynamics of computation}  can guide the optimisation of computations with respect to their energy consumption. The bounds presented here may be regarded as a contribution to such a theory.

\begin{acknowledgments}
This work is supported by  the National Natural Science Foundation of China through grant 11675136, the Hong Kong Research Grant Council through grant 17326616,  the Foundational Questions Institute through grant FQXi-RFP3-1325, the Croucher Foundation, the Swiss National Science Foundation via the National Center for Competence in Research ``QSIT" as well as via project No.\ 200020\_165843 and project No.\ 200021\_188541, and the ETH Pauli Center for Theoretical Studies. We  thank Daniel Ebler for designing the figures.

\end{acknowledgments}
 
\appendix 

\section{Proof of Theorem \ref{thm-resource}.} \label{app-mainthm}
We assume that the resource function $M$ satisfies monotonicity and regularity (Properties 1 and 3 in the main text). In addition, additivity (Property 2 in the main text) can be relaxed to:
\begin{enumerate}	 \setcounter{enumi}{1}
  	\item {\em Subadditivity on product states.} $M  (\rho \otimes \sigma) \le  M  (\rho)  +  M(\sigma)$. 
	\end{enumerate}
Under these properties, we prove a more general result on the resource requirement, which reduces to Theorem~\ref{thm-resource} (which we prove as Corollary~\ref{cor-mainthm}) when Property 2 is substituted by additivity. 

\begin{theo}\label{theo:strongest}
Every approximation  of the gate $G$ within error $\epsilon$ using a free   gate $U_G$ and a battery in the  state  $\beta$ must satisfy the inequality
\begin{align}\label{strongest}
M(\beta )  &\ge  m \overline{M}_m\left({G}\otimes G^\dag\right)-  8 \sqrt{\epsilon}K_{\rm S } m^2 -c
 \end{align}
for every $m\in\N^*$, where $\overline{M}_m(U)$ is the regularised resource generation \cite{liu2019resource,liu2019operational} of $m$ uses of a quantum gate $U$ acting on a 
system $\rm S$. Explicitly, $\overline{M}_m(U)$ is defined as
 \begin{align}\label{M-reg}
\overline{M}_m(U):=\max_{\rho_m}\frac{1}{m}\left(M\left(\map U^{\otimes m}(\rho_m)\right)-M(\rho_m)\right),
\end{align}
where $\map{U}(\cdot):=U(\cdot)U^\dag$ and the maximum is taken over all $m$-partite states.
\end{theo}    
\medskip  

We remark that Eq.\ (\ref{strongest}) is the general formula that can be used to further derive resource inequalities with simpler forms: $\overline{M}_m$ can scale differently, e.\,g.\ $\overline{M}_m=O(1)$ or $\overline{M}_m=O(m)$, for different resource theories. One can then optimise over all $m\in\N^*$ to get the scaling of the resource requirement with respect to the error, which depends on the resource theory under consideration.

\medskip

The proof of Theorem \ref{theo:strongest} is based on the following Lemma, in which  we use the notation $F_{\rm wc}  (\map C,  \map D)  :=   \inf_{\rm R} \inf_{ |\Psi\>  \in  \spc H_{\rm S}\otimes \spc H_{\rm R}}   \,   F  (   (\map C\otimes \map I_{\rm R})  (  |\Psi\>\<\Psi|), (\map D\otimes \map I_{\rm R})  (  |\Psi\>\<\Psi|)  )$. 

\begin{lemma}\label{lem:bound1}
Let $G$ be a gate acting on system $\rm S$,  $U_G$ a gate acting on system $\rm SB$,  $\beta$ be a state of system $\rm B$,   let $\map V_G$   be the channel from $\rm S$ to $\rm S B$ defined by $\map V_G  (\rho)   : =   U_G  (\rho \otimes \beta)  U_G^\dag$, and let $\map E_G$ be the channel from $\rm S$ to $\rm S$ defined by 
$\map E_G (\rho)  :  =  \Tr_{\rm B}  [\map V_G(\rho)]$. 
Then, there  exists a state $\beta'$ of system $\rm B$,  such that  
$ \|  \map V_G  -  \map G  \otimes \beta'  \|_\diamond \le 2 \sqrt{1-   F (  \map E_G,\map G)}$.  
\end{lemma}
\noindent\Proof\ Let $\widetilde\beta$ be a purification of $\beta$ with purifying system $\rm E$.   Then, the channel   $\widetilde{\map{V}}_G(\cdot):=(\map{U}_G\otimes\map{I}_{\rm E})(\cdot\otimes \widetilde \beta)$, is a Stinespring dilation of the channel $\map E_G$ \cite{paulsen2002completely}. 

The Uhlmann's theorem for gates \cite{uhlmann1976transition,kretschmann2008information}  guarantees that there exist  a Stinespring dilation of the gate $G$, say $G\otimes \widetilde \beta'$ for some pure state $\widetilde \beta'$, such that the  fidelity between  $\widetilde{\map{V}}_G$ and $G\otimes \widetilde \beta'$ is equal to the fidelity between  $\map E_G$ and $G$, namely 
\begin{align}\label{fidGUG}
F_{\rm wc}(\map{G}\otimes \widetilde \beta' ,\widetilde{\map{V}}_G )=F_{\rm wc}(\map{G} ,\map{E}_G).
\end{align}

Tracing out $\rm E$, we obtain 
\begin{align}
F_{\rm wc}(\map{G}\otimes \beta' ,\map{V}_G )\ge F_{\rm wc}(\map{G},\map{E}_G) \,,
\end{align}
where $\map{V}_G(\cdot):=\map{U}_G(\cdot\otimes \beta)$. Since  $\map{V}_G$ and $\map{G}\otimes \beta'$ are extensions of the original channels, the converse inequality also holds, namely $F_{\rm wc}(\map{G}\otimes \beta' ,\map{V}_G )\ge F_{\rm wc}(\map{G},\map{E}_G)$.   Hence,  the  inequality is in fact an equality. 

Then, the Fuchs-Van de Graph inequality \cite{fuchs1999cryptographic} yields the relation   
 \begin{align}
 \|   \map V_G      -   \map{G}\otimes \beta'     \|_\diamond     & \le    2 \sqrt{1-  F_{\rm wc}(\map{G} ,\map{E}_G)}  \, .
 \end{align}
\qed 

\medskip 
\begin{cor}\label{cor:bound2}
Let $\map V'_G$   be the channel from $\rm S$ to $\rm S B$ defined by $\map V'_G  (\rho)   : =   U_G^\dag  (\rho \otimes \beta')  U_G$. Then, one has  $ \|  \map V'_G  -  \map G^\dag  \otimes \beta  \|_\diamond \le 2 \sqrt{1-   F (  \map E_G,\map G)}$.  
\end{cor}
\noindent\Proof\  The inequality follows from the unitary invariance of the diamond norm:  
\begin{align}
\nonumber  \|  \map V'_G  -  \map G^\dag  \otimes \beta  \|_\diamond    &=   \|    \map U_G  \circ \map V'_G \circ \map G -    \map U_G \circ  (\map G^\dag  \otimes \beta  ) \circ \map G \|_\diamond \\
\nonumber &   =   \|  \map G  \otimes \beta   -  \map V_G   \|_\diamond \\
& \le 2      \sqrt{1-  F_{\rm wc}(\map{G} ,\map{E}_G)}  \, .
\end{align}
 \qed
\medskip
\begin{cor}\label{cor:4nepsilon}
Let $\map C_G$ be the multipartite  channel corresponding to the circuit in Fig.~\ref{fig:comb} of the main text. 
Then, one has the bound $ \|   \map C_G  -       ( \map G \otimes \map G^{-1})^{\otimes m}   \otimes \beta \|_\diamond  \le 4m      \sqrt{1-  F_{\rm wc}(\map{G} ,\map{E}_G)} $. 
\end{cor}
\noindent\Proof\ Follows from the unitary invariance and triangle inequality of the diamond norm, combined the bounds in Lemma \ref{lem:bound1} and Corollary \ref{cor:bound2}. \qed

\medskip  

\noindent{\bf Proof of Theorem \ref{theo:strongest}.}  Consider the input state  $\rho_{\rm in}:=\rho_{\rm in}^{(2m)}\otimes \beta $, where $\rho_{\rm in}^{(2m)}$ is an arbitrary input state on $2m$ identical copies of the system.
Let $\rho_{\rm out}$ be the output state resulting from the approximate circuit in Fig.~\ref{fig:comb} of the main text. 
Using the monotonicity of the function $M$, we obtain the relation 
\begin{align}
M(\Tr_{\rm B}  [\rho_{\rm out}])     \le M(\rho_{\rm out})    \le  M(\rho_{\rm in})  \, .
\end{align} 
 By   Property 2 (subadditivity),  we have   $M(\rho_{\rm in}) \le M(\beta)+M\left(\rho_{\rm in}^{(2m)}\right)$. Then, we have the bound
\begin{align}\label{Mbeta-inter1}
M(\beta)\ge&M (\Tr_{\rm B}[\rho_{\rm out}])-M\left(\rho_{\rm in}^{(2m)}\right) \, .
\end{align}
Now,  we apply Property 3 (regularity) to the ideal output and to its approximation, whose difference has  trace norm  at most $4m\sqrt{\epsilon}$ from the actual output state $\rho_{\rm out}$, due to Corollary \ref{cor:4nepsilon}. Noticing that the Lipschitz constant for the system  $({\rm S}\otimes {\rm S})^{\otimes m}$ is  upper bounded by $2m K_{\rm S}$, the bound (\ref{Mbeta-inter1}) becomes
\begin{align}
\nonumber M(\beta )  &\ge  M\left(\left(\map{G}\otimes\map{G}^{-1}\right)^{\otimes m}\left(\rho_{\rm in}^{(2m)}\right)\right)-M\left(\rho_{\rm in}^{(2m)}\right)&\nonumber\\
&\qquad-  8 \sqrt{\epsilon}K_{\rm S } m^2 -c
 \end{align}
 which holds for any $m\in\N^*$ and for any input state $\rho_{\rm in}^{(2m)}$. Maximising over all inputs fixing $m$, we have
\begin{align}
\nonumber M(\beta )  &\ge  m \overline{M}_m\left(G\otimes G^\dag\right)-  8 \sqrt{\epsilon}K_{\rm S } m^2 -c,
 \end{align}
 where $\overline{M}_m(U)$ is defined by Eq.\ (\ref{M-reg}).
\qed

\medskip

When $M$ is additive on product states, i.\,e.\ $M  (\rho \otimes \sigma) =  M  (\rho)  +  M(\sigma)$, the general bound (\ref{strongest}) can be simplified by finding an $m$-independent lower bound on $\overline{M}_{m}$.

\begin{cor}[Theorem \ref{thm-resource} in the main text]\label{cor-mainthm}
When $M$ is additive, the resource requirement in the battery becomes
\begin{align}\label{bound-additive}
M(\beta) &\ge\frac{(M(G)+M(G^\dag))^2}{32K_{\rm S}\sqrt{\epsilon}}-c-2K_{\rm S}\sqrt{\epsilon} \, ,
\end{align}
where $M(G):=\max_\rho M(G\rho G^\dag)-M(\rho)$ is the amount of resource generated by the gate $G$. 
\end{cor}
For additive $M$ the function $\overline{M}_m$ is monotonically increasing with $m$. Since $M(G)=\overline{M}_1(G)$, it is obviously upper bounded by $\overline{M}_m(G)$.

\noindent{\bf Proof.} 
Let us consider a product form input $\rho_{\rm in}^{(2m)}=(\rho\otimes\sigma)^{\otimes m}$ to the circuit. 
Since $M$ satisfies additivity, we have
\begin{align}
&\overline{M}_m\left(G\otimes G^\dag\right)\nonumber\\
\ge& \frac{1}{m}M\left(\left((\map{G}(\rho)\otimes\map{G}^{-1}(\sigma))^{\otimes m}\right)-M((\rho\otimes\sigma)^{\otimes m})\right)\nonumber\\
=&\left(M\left(\map{G}(\rho)\right)-M(\rho)\right)+\left(M\left(\map{G}^{-1}(\sigma)\right)-M(\sigma)\right)
\end{align}
for every $m\in\N^*$. Choosing $\rho$ and $\sigma$ to be the maximal resource generating inputs for $\map{G}$ and $\map{G}^{-1}$ respectively, we have
\begin{align}
\overline{M}_m\left(G\otimes G^\dag\right)
&\ge M(G)+M(G^\dag).
\end{align}
Substituting into Eq.\ (\ref{strongest}), we get
\begin{align}\label{app-m-final}
M(\beta )  &\ge  m \left(M(G)+M(G^\dag)\right)-  8 \sqrt{\epsilon}K_{\rm S } m^2 -c.
 \end{align}
Finally, we obtain the lower bound (\ref{bound-additive}) on the amount of resource in the battery by maximising the bound over all possible $m\in\N$. The optimal choice $m^\ast\in\N$ satisfies $|m^\ast-(M(G)+M(G^\dag))/16\sqrt{\epsilon}K_{\rm S}|\le 1/2$. Substituting into Eq.~(\ref{app-m-final}) we get (\ref{bound-additive}).
\qed

We conclude by mentioning a further extension of Theorem \ref{theo:strongest} that takes into account the possibility of applying the gate $ G$ on part of a composite system:  

\begin{cor}
Every approximation  of the gate $G$ within error $\epsilon$ using a free   gate $U_G$ and a battery in the  state  $\beta$ must satisfy the inequality
\begin{align}\label{strongest2}
M(\beta )  &\ge m \overline{M}_m\left(G\otimes I_{\rm R}\otimes G^\dag\otimes I_{\rm R}\right)-  8 \sqrt{\epsilon}K_{\rm SR} m^2 -c
 \end{align}
for every $m\in\N^*$
where $\rm R$ is a reference system. 
\end{cor} 

\noindent\Proof\ The result follows from the application of Theorem \ref{theo:strongest} to the gate $G\otimes  I_{\rm R}$, observing that, by definition, the diamond norm and the worst-case fidelity are invariant under addition of a reference system. \qed

\section{Application to the resource theory of coherence} \label{app-coherence}
The resource of quantum coherence \cite{baumgratz2014quantifying,yadin2016quantum,winter2016operational,chitambar2016critical,chitambar2016comparison,marvian2016quantify} can be characterised operationally in terms of different sets of free operations, such as    strictly incoherent operations \cite{yadin2016quantum}, maximally incoherent operations \cite{aberg2006quantifying,diaz2018using}, dephasing covariant operations \cite{chitambar2016critical,chitambar2016comparison,marvian2016quantify}, phase covariant operations \cite{marvian2016quantify}, and physically incoherent operations \cite{chitambar2016critical,chitambar2016comparison}.  
These operations are defined relative to a fixed basis $\{  |i\>\}$, and preserve the set of incoherent states, of the form $\rho  =  \sum_i  \, p_i \,|i\>\<i|$.  For composite systems, it is understood that the fixed basis of the composite system is the product of the fixed bases for the components. 

For the purpose of our bound, the choice of the set of free operations is not critical. 
As a  measure of resource, we consider the relative entropy of coherence 
 \cite{baumgratz2014quantifying}
\begin{align}\label{coherence-measure}
C(\rho):=S(\rho_{\rm diag})-S(\rho) \, ,
\end{align}
$S$ denoting the von Neumann entropy of quantum states and $\rho_{\rm diag}$ being the diagonal part of $\rho$ in the energy basis.
This measure of coherence satisfies the Properties 1 (Monotonicity) and 2 (Additivity on product states). It also satisfies Property 3, as shown by the following 

\begin{prop}
The function  $ C:   L(\C^d)  \to \R \, ,   C(\rho)=S(\rho_{\rm diag})-S(\rho)$ satisfies the inequality    $|    \map C (\rho)   -  \map C (\sigma)  |  \le \log d  \,  \|\rho - \sigma\|_1   + 2$. 
\end{prop}

\noindent\Proof\ For any two states $\rho$ and $\sigma$ in a $d$-dimensional Hilbert space, the difference of their entropies is bounded by the Fannes-Audenaert inequality \cite{fannes1973continuity,audenaert2007sharp}
\begin{align}
|S(\rho)-S(\sigma)|\le\frac{\log d}{2}\|\rho-\sigma\|_1+h_2(\|\rho-\sigma\|_1/2).
\end{align}
where $h_2(p):=-p\log p-(1-p)\log(1-p)$ is the binary entropy, upper bounded by one for any $p$. For our purpose, it is enough to use the relaxed version of the above inequality:
\begin{align}\label{fannes}
|S(\rho)-S(\sigma)|\le\frac{\log d}{2}\|\rho-\sigma\|_1+1.
\end{align}

Now, let us consider the difference of the relative entropies of coherence (\ref{coherence-measure}) between $\rho$ and $\sigma$. We have
\begin{align}
|C(\rho)-C(\sigma)|\le |S(\rho)-S(\sigma)|+|S(\rho_{\rm diag})-S(\sigma_{\rm diag})|.
\end{align}
Applying Eq.\ (\ref{fannes}) to both terms on the right hand side of the above inequality and noticing that $\|\rho_{\rm diag}-\sigma_{\rm diag}\|_1\le\|\rho-\sigma\|_1$ (monotonicity of trace distance under data processing), we have
\begin{align}
|C(\rho)-C(\sigma)|\le \log d\cdot\|\rho-\sigma\|_1+2.
\end{align}
Therefore, we have $K=\log d$ and $c=2$.\qed 

\medskip  

Using the above Proposition and  Theorem \ref{thm-resource} of the main text, we obtain a lower bound on the initial coherence in the battery:
\begin{align}
C(\beta)\ge\frac{\left(C(G)+C(G^\dag)\right)^2}{32\sqrt{\epsilon}\log d_{\rm S}}-2.
\end{align}
Gates like the generalized Hadamard gate have coherence generation up to $\log d_{\rm S}$. Therefore, the minimum amount of required coherence in a quantum processor is lower bounded as
\begin{align}\label{lower-bound-coherence}
C(|\beta\>\<\beta|)\ge\frac{\log d_{\rm S}}{8\sqrt{\epsilon}}-2 \, .
\end{align}

\section{Lower bound on the accuracy}\label{app-lowerbound}
In the following we will  determine the lower bound (\ref{Fwc}) on $F_{\rm wc}$. Notice that $F_{\rm wc}$ can be rewritten as $F_{\rm wc}=\inf_{\rm R}\inf_{|\Psi\>  \in  \spc H_{\rm S} \otimes \spc H_{\rm R}} F_{\Psi}$,
where
\begin{align}\label{fidelity}
F_{\Psi}:= \Tr\left[(\map E_{G} \otimes \map I_{\rm R})  (  \Psi ) \,   (\map{G} \otimes \map{I}_{\rm R})(\Psi)   \right]\,.
\end{align}
To evaluate the this fidelity, we observe  that the gate $U_G$, defined in Eq.~(\ref{ug}), can be expressed as $  U_G  =  U_G^{(\rm ok)}  +    P^{(\rm ok)}_{ \perp}$, where $P^{(\rm ok)}_{\perp}$ is the projector on the eigenstates of the total energy outside the set $\set E_{\rm ok}$, and  $ U_G^{(\rm ok)}$ is the partial isometry 
\begin{align}\label{uok}
U^{(\rm ok)}_G  &:=   \sum_{x,y}   G_{xy}  \,  |\psi_x\>\<\psi_y|  \otimes S^{(xy)}  \\
S^{(xy)}  &:=\sum_{E\in\set{E}_{\rm ok}}  \,  |E  -   E_{{\rm S}, x}\>\<  E-  E_{{\rm S}, y}|   \, ,
\end{align}
where we used the shorthand $A_{xy}  =  \<\psi_x|  A  |\psi_y\>$ for a generic operator $ A  \in  L(  \spc H_{\rm S})$. 
Observe that the battery state (\ref{sine}) is defined so that the joint state of the system and the battery has full support in energy subspaces with $E\in\set{E}_{\rm ok}$.
Substituting (\ref{uok}) into   (\ref{fidelity}), one has  the expression 
\begin{align}\label{fidelitysum}
F_{\Psi}   =\sum_{x,y,z,t  = 0}^{d_{\rm S}-1} C_{xyzt}(\rho G^\dag )_{xy}G_{yx}G^\dag_{zt} (G \rho)_{tz},       
\end{align}
where $\rho$ is the marginal state  $\rho= \Tr_{\rm R} [|\Psi\>\<\Psi|] $  and  $C_{xyzt}   =  \<\beta| \,  S^\dag_{zt} S_{xy} \, |\beta\>$.   

The quantity $C_{xyzt}$  can be explicitly evaluated as 
\begin{align}
C_{xyzt}  = &\sum_{k=\|H_{\rm S}\|+y}^{(R-1)\|H_{\rm S}\|-x}\frac{2}{L}\sin\left(\frac{(k-z+t-2\|H_{\rm S}\|+1)\pi}{L}\right)&\nonumber\\
&\qquad\times \sin\left(\frac{(k+x-y-2\|H_{\rm S}\|+1)\pi}{L}\right)\nonumber\\
\nonumber =&\frac{(L-x-y-1)\cos\left(\frac{(x-y-z+t)\pi}{L}\right)}{L}\\
&\qquad\nonumber+\frac{\sin\left(\frac{(x+y+1)\pi}{L}\right)\cos\left(\frac{(2\|H_{\rm S}\|-t+z)\pi}{L}\right)}{L\sin\left(\frac{\pi}{L}\right)}\\
=&1-  \frac{(x-y-z+t)^2  \pi^2}{ 8\< H_{\rm B}  \>^2}\left(1+O\left(\frac{\|H_{\rm S}\|}{\<  H_{\rm B}  \>}\right)\right) \,,\label{C-xyzt}
\end{align} 
where the last step follows from the definition of $L$.  
Inserting the above expression into Eq.\ (\ref{fidelitysum}) and rearranging the different terms, we obtain
\begin{align}\label{feta}
F_{\Psi}     =  1-\frac{\pi^2 \,    \Var (  \Delta_G H_{\rm S} )  }{  4  \<  H_{\rm B}  \>^2}  \,      \left(1+O\left(\frac{\|H_{\rm S}\|}{\< H_{\rm B }\>}\right)\right)   \, , 
\end{align}
where $\Var (\Delta_G H_{\rm S})$ denotes the variance of the operator $\Delta_G H_{\rm S}$ on the state $|\Psi\>$. 
Noting that $\Var(\Delta_G H_{\rm S} ) \le \left((\lambda_{\max}-\lambda_{\min})(\Delta_G H_{\rm S})/2\right)^2 $, Eq.\ (\ref{feta}) implies the following bound on the worst-case fidelity
\begin{align}
F_{\rm wc}     \ge 1-\left(\frac{\pi \left(\lambda_{\max}-\lambda_{\min}\right)(\Delta_G H_{\rm S})}{ 4  \<  H_{\rm B}  \>}\right)^2  \,      \left(1+O\left(\frac{\|H_{\rm S}\|}{\< H_{\rm B }\>}\right)\right)   \, .
\end{align}

\section{Energy requirement in terms of the diamond norm error}\label{app-diamond}
Here we show that the energy requirement still scales as $1/\sqrt{\epsilon}$, when the error $\epsilon$ is measured by the diamond norm error \cite{diamondnote} instead of $1-F_{\rm wc}$. 

On one hand, since the diamond norm error upper bounds the worst-case infidelity via the inequality $1-\sqrt{F_{\rm wc}}\le \frac{1}{2}\|\map{E}_G-\map{G}\|_{\diamond}$, we have $1-F_{\rm wc}\le 2\epsilon$ when the diamond norm error is at most $\epsilon$. The proof of the lower bound [cf.~Eq.~(\ref{lower})] goes through and we get 
\begin{align}\label{diamond-lower}
\<  H_{\rm B}  \>    \ge     \frac{\| H_{\rm S}\| }{  8 \sqrt {2\epsilon}    }   - O(\sqrt{\epsilon}) \, . 
\end{align}

On the other hand, the output state of the construction (\ref{ug}) can be expressed as
\begin{align}
\map{E}_G\otimes\map{I}_{\rm R}(\Psi)&=\sum_{x,y,z,t=0}^{d_{\rm S}-1}C_{xyzt}G_{xy}|\psi_x\>\<\psi_y|\Psi|\psi_t\>\<\psi_z|G^\ast_{zt}
\end{align}
where $C_{xyzt}$ is given by Eq.~(\ref{C-xyzt}). The diamond norm error can be obtained by taking the worst case over $\Psi$ of the following quantity:
\begin{align}\label{app-diamond-temp}
\frac{1}{2}\left\|\sum_{x,y,z,t=0}^{d_{\rm S}-1}(C_{xyzt}-1)G_{xy}|\psi_x\>\<\psi_y|\Psi|\psi_t\>\<\psi_z|G^\ast_{zt}\right\|_1.
\end{align}
The quantity $1-C_{xyzt}$ can be upper bounded as 
\begin{align}
1-C_{xyzt}\le\frac{\pi^2(d_{\rm S}-1)^2}{2\<H_{\rm B}\>^2}\left(1+O\left(\frac{\|H_{\rm S}\|}{\<  H_{\rm B}  \>}\right)\right)\quad\forall\,x,y,z,t,
\end{align}
since $x,y,z,t\in\{0,\dots,d_{\rm S}-1\}$. Substituting into Eq.~(\ref{app-diamond-temp}), the diamond norm error $\epsilon$ can be upper bounded as
\begin{align}
\epsilon&\le \frac{\pi^2(d_{\rm S}-1)^2}{2\<H_{\rm B}\>^2}\left(1+O\left(\frac{\|H_{\rm S}\|}{\<  H_{\rm B}  \>}\right)\right)\max_{\Psi}\|G\Psi G^\dag\|_1\nonumber\\
&=\frac{\pi^2(d_{\rm S}-1)^2}{2\<H_{\rm B}\>^2}\left(1+O\left(\frac{\|H_{\rm S}\|}{\<  H_{\rm B}  \>}\right)\right).
\end{align}
Finally, we have
\begin{align}
\<H_{\rm B}\>\le\frac{\pi(d_{\rm S}-1)}{\sqrt{2\epsilon}}\left(1+O\left(\frac{\|H_{\rm S}\|}{\<  H_{\rm B}  \>}\right)\right).
\end{align}
In summary, we derived both upper and lower bounds on $\<H_{\rm B}\>$ in terms of the diamond norm error. Since the two bounds have matching scaling, we conclude that the energy requirement scales as $1/\sqrt{\epsilon}$, independently of whether one measures the error in terms of the worst-case infidelity or in terms of the diamond norm error.

\section{Perfect implementation of quantum computation} \label{app-computing}
Here we consider a generic quantum algorithm that starts by preparing an energy eigenstate state $|\psi_x\>$ and ends by measuring the energy eigenbasis. The overall action of the algorithm can be described by a unitary gate $G$.  We  observe that the input-output relation induced by gate $G$ can be reproduced without errors   using the interaction (\ref{ug}).
For an initial state $|\psi_x\>$ of the system, one prepares the battery in the state $|E-E_{{\rm S}, x}\>$, so that the joint state is
\begin{align}
|x,  E\>    =    |\psi_x\>  \otimes  |E-  E_{{\rm S}, x}\>,
\end{align}
where the total energy $E\in\set{E}_{\rm ok}$. 
Then the initial state of the system and the battery can be expressed as $|x, E\>$. The effect of the interaction (\ref{ug}) can be expressed as $U_G|x, E\>=\sum_y g_{y,x}|y,E\>$, where $g_{x,y}=\<\psi_y|G|\psi_x\>$ is the matrix element of $G$. The system ends up in the state $\sum_y |g_{y,x}|^2|\psi_y\>\<\psi_y|$. Therefore, when measuring in the energy eigenbasis in the end, the probability of getting the outcome $y$ is exactly $|g_{y,x}|^2$, which is the same as the original algorithm $G$.


\begin{thebibliography}{51}%
\makeatletter
\providecommand \@ifxundefined [1]{%
 \@ifx{#1\undefined}
}%
\providecommand \@ifnum [1]{%
 \ifnum #1\expandafter \@firstoftwo
 \else \expandafter \@secondoftwo
 \fi
}%
\providecommand \@ifx [1]{%
 \ifx #1\expandafter \@firstoftwo
 \else \expandafter \@secondoftwo
 \fi
}%
\providecommand \natexlab [1]{#1}%
\providecommand \enquote  [1]{``#1''}%
\providecommand \bibnamefont  [1]{#1}%
\providecommand \bibfnamefont [1]{#1}%
\providecommand \citenamefont [1]{#1}%
\providecommand \href@noop [0]{\@secondoftwo}%
\providecommand \href [0]{\begingroup \@sanitize@url \@href}%
\providecommand \@href[1]{\@@startlink{#1}\@@href}%
\providecommand \@@href[1]{\endgroup#1\@@endlink}%
\providecommand \@sanitize@url [0]{\catcode `\\12\catcode `\$12\catcode
  `\&12\catcode `\#12\catcode `\^12\catcode `\_12\catcode `\%12\relax}%
\providecommand \@@startlink[1]{}%
\providecommand \@@endlink[0]{}%
\providecommand \url  [0]{\begingroup\@sanitize@url \@url }%
\providecommand \@url [1]{\endgroup\@href {#1}{\urlprefix }}%
\providecommand \urlprefix  [0]{URL }%
\providecommand \Eprint [0]{\href }%
\providecommand \doibase [0]{https://doi.org/}%
\providecommand \selectlanguage [0]{\@gobble}%
\providecommand \bibinfo  [0]{\@secondoftwo}%
\providecommand \bibfield  [0]{\@secondoftwo}%
\providecommand \translation [1]{[#1]}%
\providecommand \BibitemOpen [0]{}%
\providecommand \bibitemStop [0]{}%
\providecommand \bibitemNoStop [0]{.\EOS\space}%
\providecommand \EOS [0]{\spacefactor3000\relax}%
\providecommand \BibitemShut  [1]{\csname bibitem#1\endcsname}%
\let\auto@bib@innerbib\@empty
\bibitem [{\citenamefont {Landauer}(1961)}]{landauer1961irreversibility}%
  \BibitemOpen
  \bibfield  {author} {\bibinfo {author} {\bibfnamefont {R.}~\bibnamefont
  {Landauer}},\ }\bibfield  {title} {\bibinfo {title} {Irreversibility and heat
  generation in the computing process},\ }\href@noop {} {\bibfield  {journal}
  {\bibinfo  {journal} {IBM journal of Research and Development}\ }\textbf
  {\bibinfo {volume} {5}},\ \bibinfo {pages} {183} (\bibinfo {year}
  {1961})}\BibitemShut {NoStop}%
\bibitem [{\citenamefont {Nielsen}\ and\ \citenamefont
  {Chuang}(2000)}]{nielsen2000quantum}%
  \BibitemOpen
  \bibfield  {author} {\bibinfo {author} {\bibfnamefont {M.~A.}\ \bibnamefont
  {Nielsen}}\ and\ \bibinfo {author} {\bibfnamefont {I.}~\bibnamefont
  {Chuang}},\ }\bibfield  {title} {\bibinfo {title} {Quantum computation},\
  }\href@noop {} {\bibfield  {journal} {\bibinfo  {journal} {Quantum
  Information. Cambridge University Press, Cambridge}\ } (\bibinfo {year}
  {2000})}\BibitemShut {NoStop}%
\bibitem [{\citenamefont {P{\'e}rez-Garc{\'\i}a}(2006)}]{perez2006optimality}%
  \BibitemOpen
  \bibfield  {author} {\bibinfo {author} {\bibfnamefont {D.}~\bibnamefont
  {P{\'e}rez-Garc{\'\i}a}},\ }\bibfield  {title} {\bibinfo {title} {Optimality
  of programmable quantum measurements},\ }\href@noop {} {\bibfield  {journal}
  {\bibinfo  {journal} {Physical Review A}\ }\textbf {\bibinfo {volume} {73}},\
  \bibinfo {pages} {052315} (\bibinfo {year} {2006})}\BibitemShut {NoStop}%
\bibitem [{\citenamefont {Kubicki}\ \emph {et~al.}(2019)\citenamefont
  {Kubicki}, \citenamefont {Palazuelos},\ and\ \citenamefont
  {P{\'e}rez-Garc{\'\i}a}}]{kubicki2019resource}%
  \BibitemOpen
  \bibfield  {author} {\bibinfo {author} {\bibfnamefont {A.~M.}\ \bibnamefont
  {Kubicki}}, \bibinfo {author} {\bibfnamefont {C.}~\bibnamefont
  {Palazuelos}},\ and\ \bibinfo {author} {\bibfnamefont {D.}~\bibnamefont
  {P{\'e}rez-Garc{\'\i}a}},\ }\bibfield  {title} {\bibinfo {title} {Resource
  quantification for the no-programing theorem},\ }\href@noop {} {\bibfield
  {journal} {\bibinfo  {journal} {Physical Review Letters}\ }\textbf {\bibinfo
  {volume} {122}},\ \bibinfo {pages} {080505} (\bibinfo {year}
  {2019})}\BibitemShut {NoStop}%
\bibitem [{\citenamefont {Yang}\ \emph {et~al.}(2020)\citenamefont {Yang},
  \citenamefont {Renner},\ and\ \citenamefont {Chiribella}}]{yang2020optimal}%
  \BibitemOpen
  \bibfield  {author} {\bibinfo {author} {\bibfnamefont {Y.}~\bibnamefont
  {Yang}}, \bibinfo {author} {\bibfnamefont {R.}~\bibnamefont {Renner}},\ and\
  \bibinfo {author} {\bibfnamefont {G.}~\bibnamefont {Chiribella}},\ }\bibfield
   {title} {\bibinfo {title} {Optimal universal programming of unitary gates},\
  }\href@noop {} {\bibfield  {journal} {\bibinfo  {journal} {Physical Review
  Letters}\ }\textbf {\bibinfo {volume} {125}},\ \bibinfo {pages} {210501}
  (\bibinfo {year} {2020})}\BibitemShut {NoStop}%
\bibitem [{\citenamefont {Sagawa}\ and\ \citenamefont
  {Ueda}(2009)}]{sagawa2009minimal}%
  \BibitemOpen
  \bibfield  {author} {\bibinfo {author} {\bibfnamefont {T.}~\bibnamefont
  {Sagawa}}\ and\ \bibinfo {author} {\bibfnamefont {M.}~\bibnamefont {Ueda}},\
  }\bibfield  {title} {\bibinfo {title} {Minimal energy cost for thermodynamic
  information processing: measurement and information erasure},\ }\href@noop {}
  {\bibfield  {journal} {\bibinfo  {journal} {Physical Review Letters}\
  }\textbf {\bibinfo {volume} {102}},\ \bibinfo {pages} {250602} (\bibinfo
  {year} {2009})}\BibitemShut {NoStop}%
\bibitem [{\citenamefont {Reeb}\ and\ \citenamefont
  {Wolf}(2014)}]{reeb2014improved}%
  \BibitemOpen
  \bibfield  {author} {\bibinfo {author} {\bibfnamefont {D.}~\bibnamefont
  {Reeb}}\ and\ \bibinfo {author} {\bibfnamefont {M.~M.}\ \bibnamefont
  {Wolf}},\ }\bibfield  {title} {\bibinfo {title} {An improved landauer
  principle with finite-size corrections},\ }\href@noop {} {\bibfield
  {journal} {\bibinfo  {journal} {New Journal of Physics}\ }\textbf {\bibinfo
  {volume} {16}},\ \bibinfo {pages} {103011} (\bibinfo {year}
  {2014})}\BibitemShut {NoStop}%
\bibitem [{\citenamefont {Faist}\ \emph {et~al.}(2015)\citenamefont {Faist},
  \citenamefont {Dupuis}, \citenamefont {Oppenheim},\ and\ \citenamefont
  {Renner}}]{faist2015minimal}%
  \BibitemOpen
  \bibfield  {author} {\bibinfo {author} {\bibfnamefont {P.}~\bibnamefont
  {Faist}}, \bibinfo {author} {\bibfnamefont {F.}~\bibnamefont {Dupuis}},
  \bibinfo {author} {\bibfnamefont {J.}~\bibnamefont {Oppenheim}},\ and\
  \bibinfo {author} {\bibfnamefont {R.}~\bibnamefont {Renner}},\ }\bibfield
  {title} {\bibinfo {title} {The minimal work cost of information processing},\
  }\href@noop {} {\bibfield  {journal} {\bibinfo  {journal} {Nature
  Communications}\ }\textbf {\bibinfo {volume} {6}} (\bibinfo {year}
  {2015})}\BibitemShut {NoStop}%
\bibitem [{\citenamefont {Faist}\ and\ \citenamefont
  {Renner}(2018)}]{faist2018fundamental}%
  \BibitemOpen
  \bibfield  {author} {\bibinfo {author} {\bibfnamefont {P.}~\bibnamefont
  {Faist}}\ and\ \bibinfo {author} {\bibfnamefont {R.}~\bibnamefont {Renner}},\
  }\bibfield  {title} {\bibinfo {title} {Fundamental work cost of quantum
  processes},\ }\href@noop {} {\bibfield  {journal} {\bibinfo  {journal}
  {Physical Review X}\ }\textbf {\bibinfo {volume} {8}},\ \bibinfo {pages}
  {021011} (\bibinfo {year} {2018})}\BibitemShut {NoStop}%
\bibitem [{\citenamefont {Wigner}(1952)}]{wigner1952messung}%
  \BibitemOpen
  \bibfield  {author} {\bibinfo {author} {\bibfnamefont {E.}~\bibnamefont
  {Wigner}},\ }\bibfield  {title} {\bibinfo {title} {Die messung
  quantenmechanischer operatoren},\ }\href@noop {} {\bibfield  {journal}
  {\bibinfo  {journal} {Zeitschrift f{\"u}r Physik A Hadrons and Nuclei}\
  }\textbf {\bibinfo {volume} {133}},\ \bibinfo {pages} {101} (\bibinfo {year}
  {1952})}\BibitemShut {NoStop}%
\bibitem [{\citenamefont {Araki}\ and\ \citenamefont
  {Yanase}(1960)}]{araki1960measurement}%
  \BibitemOpen
  \bibfield  {author} {\bibinfo {author} {\bibfnamefont {H.}~\bibnamefont
  {Araki}}\ and\ \bibinfo {author} {\bibfnamefont {M.~M.}\ \bibnamefont
  {Yanase}},\ }\bibfield  {title} {\bibinfo {title} {Measurement of quantum
  mechanical operators},\ }\href@noop {} {\bibfield  {journal} {\bibinfo
  {journal} {Physical Review}\ }\textbf {\bibinfo {volume} {120}},\ \bibinfo
  {pages} {622} (\bibinfo {year} {1960})}\BibitemShut {NoStop}%
\bibitem [{\citenamefont {Ozawa}(2002)}]{ozawa2002conservative}%
  \BibitemOpen
  \bibfield  {author} {\bibinfo {author} {\bibfnamefont {M.}~\bibnamefont
  {Ozawa}},\ }\bibfield  {title} {\bibinfo {title} {Conservative quantum
  computing},\ }\href@noop {} {\bibfield  {journal} {\bibinfo  {journal}
  {Physical Review Letters}\ }\textbf {\bibinfo {volume} {89}},\ \bibinfo
  {pages} {057902} (\bibinfo {year} {2002})}\BibitemShut {NoStop}%
\bibitem [{\citenamefont {Ozawa}(2003)}]{ozawa2003uncertainty}%
  \BibitemOpen
  \bibfield  {author} {\bibinfo {author} {\bibfnamefont {M.}~\bibnamefont
  {Ozawa}},\ }\bibfield  {title} {\bibinfo {title} {Uncertainty principle for
  quantum instruments and computing},\ }\href@noop {} {\bibfield  {journal}
  {\bibinfo  {journal} {International Journal of Quantum Information}\ }\textbf
  {\bibinfo {volume} {1}},\ \bibinfo {pages} {569} (\bibinfo {year}
  {2003})}\BibitemShut {NoStop}%
\bibitem [{\citenamefont {Gea-Banacloche}\ and\ \citenamefont
  {Ozawa}(2006)}]{gea2006minimum}%
  \BibitemOpen
  \bibfield  {author} {\bibinfo {author} {\bibfnamefont {J.}~\bibnamefont
  {Gea-Banacloche}}\ and\ \bibinfo {author} {\bibfnamefont {M.}~\bibnamefont
  {Ozawa}},\ }\bibfield  {title} {\bibinfo {title} {Minimum-energy pulses for
  quantum logic cannot be shared},\ }\href@noop {} {\bibfield  {journal}
  {\bibinfo  {journal} {Physical Review A}\ }\textbf {\bibinfo {volume} {74}},\
  \bibinfo {pages} {060301} (\bibinfo {year} {2006})}\BibitemShut {NoStop}%
\bibitem [{\citenamefont {Karasawa}\ and\ \citenamefont
  {Ozawa}(2007)}]{karasawa2007conservation}%
  \BibitemOpen
  \bibfield  {author} {\bibinfo {author} {\bibfnamefont {T.}~\bibnamefont
  {Karasawa}}\ and\ \bibinfo {author} {\bibfnamefont {M.}~\bibnamefont
  {Ozawa}},\ }\bibfield  {title} {\bibinfo {title} {Conservation-law-induced
  quantum limits for physical realizations of the quantum not gate},\
  }\href@noop {} {\bibfield  {journal} {\bibinfo  {journal} {Physical Review
  A}\ }\textbf {\bibinfo {volume} {75}},\ \bibinfo {pages} {032324} (\bibinfo
  {year} {2007})}\BibitemShut {NoStop}%
\bibitem [{\citenamefont {Karasawa}\ \emph {et~al.}(2009)\citenamefont
  {Karasawa}, \citenamefont {Gea-Banacloche},\ and\ \citenamefont
  {Ozawa}}]{karasawa2009gate}%
  \BibitemOpen
  \bibfield  {author} {\bibinfo {author} {\bibfnamefont {T.}~\bibnamefont
  {Karasawa}}, \bibinfo {author} {\bibfnamefont {J.}~\bibnamefont
  {Gea-Banacloche}},\ and\ \bibinfo {author} {\bibfnamefont {M.}~\bibnamefont
  {Ozawa}},\ }\bibfield  {title} {\bibinfo {title} {Gate fidelity of arbitrary
  single-qubit gates constrained by conservation laws},\ }\href@noop {}
  {\bibfield  {journal} {\bibinfo  {journal} {Journal of Physics A:
  Mathematical and Theoretical}\ }\textbf {\bibinfo {volume} {42}},\ \bibinfo
  {pages} {225303} (\bibinfo {year} {2009})}\BibitemShut {NoStop}%
\bibitem [{\citenamefont {Tajima}\ \emph {et~al.}(2018)\citenamefont {Tajima},
  \citenamefont {Shiraishi},\ and\ \citenamefont
  {Saito}}]{tajima2018uncertainty}%
  \BibitemOpen
  \bibfield  {author} {\bibinfo {author} {\bibfnamefont {H.}~\bibnamefont
  {Tajima}}, \bibinfo {author} {\bibfnamefont {N.}~\bibnamefont {Shiraishi}},\
  and\ \bibinfo {author} {\bibfnamefont {K.}~\bibnamefont {Saito}},\ }\bibfield
   {title} {\bibinfo {title} {Uncertainty relations in implementation of
  unitary operations},\ }\href@noop {} {\bibfield  {journal} {\bibinfo
  {journal} {Physical Review Letters}\ }\textbf {\bibinfo {volume} {121}},\
  \bibinfo {pages} {110403} (\bibinfo {year} {2018})}\BibitemShut {NoStop}%
\bibitem [{\citenamefont {Fredkin}\ and\ \citenamefont
  {Toffoli}(1982)}]{fredkin1982conservative}%
  \BibitemOpen
  \bibfield  {author} {\bibinfo {author} {\bibfnamefont {E.}~\bibnamefont
  {Fredkin}}\ and\ \bibinfo {author} {\bibfnamefont {T.}~\bibnamefont
  {Toffoli}},\ }\bibfield  {title} {\bibinfo {title} {Conservative logic},\
  }\href@noop {} {\bibfield  {journal} {\bibinfo  {journal} {International
  Journal of Theoretical Physics}\ }\textbf {\bibinfo {volume} {21}},\ \bibinfo
  {pages} {219} (\bibinfo {year} {1982})}\BibitemShut {NoStop}%
\bibitem [{\citenamefont {Ikonen}\ \emph {et~al.}(2017)\citenamefont {Ikonen},
  \citenamefont {Salmilehto},\ and\ \citenamefont
  {M{\"o}tt{\"o}nen}}]{ikonen2017energy}%
  \BibitemOpen
  \bibfield  {author} {\bibinfo {author} {\bibfnamefont {J.}~\bibnamefont
  {Ikonen}}, \bibinfo {author} {\bibfnamefont {J.}~\bibnamefont {Salmilehto}},\
  and\ \bibinfo {author} {\bibfnamefont {M.}~\bibnamefont {M{\"o}tt{\"o}nen}},\
  }\bibfield  {title} {\bibinfo {title} {Energy-efficient quantum computing},\
  }\href@noop {} {\bibfield  {journal} {\bibinfo  {journal} {npj Quantum
  Information}\ }\textbf {\bibinfo {volume} {3}},\ \bibinfo {pages} {17}
  (\bibinfo {year} {2017})}\BibitemShut {NoStop}%
\bibitem [{\citenamefont {Coecke}\ \emph {et~al.}(2016)\citenamefont {Coecke},
  \citenamefont {Fritz},\ and\ \citenamefont
  {Spekkens}}]{coecke2016mathematical}%
  \BibitemOpen
  \bibfield  {author} {\bibinfo {author} {\bibfnamefont {B.}~\bibnamefont
  {Coecke}}, \bibinfo {author} {\bibfnamefont {T.}~\bibnamefont {Fritz}},\ and\
  \bibinfo {author} {\bibfnamefont {R.~W.}\ \bibnamefont {Spekkens}},\
  }\bibfield  {title} {\bibinfo {title} {A mathematical theory of resources},\
  }\href@noop {} {\bibfield  {journal} {\bibinfo  {journal} {Information and
  Computation}\ }\textbf {\bibinfo {volume} {250}},\ \bibinfo {pages} {59}
  (\bibinfo {year} {2016})}\BibitemShut {NoStop}%
\bibitem [{\citenamefont {Kitaev}(1997)}]{kitaev1997quantum}%
  \BibitemOpen
  \bibfield  {author} {\bibinfo {author} {\bibfnamefont {A.~Y.}\ \bibnamefont
  {Kitaev}},\ }\bibfield  {title} {\bibinfo {title} {Quantum computations:
  algorithms and error correction},\ }\href@noop {} {\bibfield  {journal}
  {\bibinfo  {journal} {Russian Mathematical Surveys}\ }\textbf {\bibinfo
  {volume} {52}},\ \bibinfo {pages} {1191} (\bibinfo {year}
  {1997})}\BibitemShut {NoStop}%
\bibitem [{dia()}]{diamondnote}%
  \BibitemOpen
  \href@noop {} {}\bibinfo {note} {The diamond norm is defined as the maximum
  trace distance between the outputs of the two channels, maximized over all
  input states and over all possible reference
  systems~\cite{kitaev1997quantum}.}\BibitemShut {Stop}%
\bibitem [{\citenamefont {Kretschmann}\ \emph
  {et~al.}(2008{\natexlab{a}})\citenamefont {Kretschmann}, \citenamefont
  {Schlingemann},\ and\ \citenamefont {Werner}}]{kretschmann2008information}%
  \BibitemOpen
  \bibfield  {author} {\bibinfo {author} {\bibfnamefont {D.}~\bibnamefont
  {Kretschmann}}, \bibinfo {author} {\bibfnamefont {D.}~\bibnamefont
  {Schlingemann}},\ and\ \bibinfo {author} {\bibfnamefont {R.~F.}\ \bibnamefont
  {Werner}},\ }\bibfield  {title} {\bibinfo {title} {The
  information-disturbance tradeoff and the continuity of stinespring's
  representation},\ }\href@noop {} {\bibfield  {journal} {\bibinfo  {journal}
  {IEEE transactions on Information Theory}\ }\textbf {\bibinfo {volume}
  {54}},\ \bibinfo {pages} {1708} (\bibinfo {year}
  {2008}{\natexlab{a}})}\BibitemShut {NoStop}%
\bibitem [{\citenamefont {Chiribella}\ \emph {et~al.}(2013)\citenamefont
  {Chiribella}, \citenamefont {D'Ariano}, \citenamefont {Perinotti},
  \citenamefont {Schlingemann},\ and\ \citenamefont
  {Werner}}]{chiribella2013short}%
  \BibitemOpen
  \bibfield  {author} {\bibinfo {author} {\bibfnamefont {G.}~\bibnamefont
  {Chiribella}}, \bibinfo {author} {\bibfnamefont {G.~M.}\ \bibnamefont
  {D'Ariano}}, \bibinfo {author} {\bibfnamefont {P.}~\bibnamefont {Perinotti}},
  \bibinfo {author} {\bibfnamefont {D.}~\bibnamefont {Schlingemann}},\ and\
  \bibinfo {author} {\bibfnamefont {R.}~\bibnamefont {Werner}},\ }\bibfield
  {title} {\bibinfo {title} {A short impossibility proof of quantum bit
  commitment},\ }\href@noop {} {\bibfield  {journal} {\bibinfo  {journal}
  {Physics Letters A}\ }\textbf {\bibinfo {volume} {377}},\ \bibinfo {pages}
  {1076} (\bibinfo {year} {2013})}\BibitemShut {NoStop}%
\bibitem [{\citenamefont {Gutoski}\ \emph {et~al.}(2017)\citenamefont
  {Gutoski}, \citenamefont {Rosmanis},\ and\ \citenamefont
  {Sikora}}]{gutoski2017fidelity}%
  \BibitemOpen
  \bibfield  {author} {\bibinfo {author} {\bibfnamefont {G.}~\bibnamefont
  {Gutoski}}, \bibinfo {author} {\bibfnamefont {A.}~\bibnamefont {Rosmanis}},\
  and\ \bibinfo {author} {\bibfnamefont {J.}~\bibnamefont {Sikora}},\
  }\bibfield  {title} {\bibinfo {title} {Fidelity of quantum strategies with
  applications to cryptography},\ }\href@noop {} {\bibfield  {journal}
  {\bibinfo  {journal} {arXiv preprint arXiv:1704.04033}\ } (\bibinfo {year}
  {2017})}\BibitemShut {NoStop}%
\bibitem [{2fo()}]{2footnote}%
  \BibitemOpen
  \href@noop {} {}\bibinfo {note} {Similar techniques have been used in Refs.\
  \cite{kretschmann2008complementarity,tajima2018uncertainty}.}\BibitemShut
  {Stop}%
\bibitem [{\citenamefont {Chiribella}\ and\ \citenamefont
  {Yang}(2017)}]{chiribella2017optimal}%
  \BibitemOpen
  \bibfield  {author} {\bibinfo {author} {\bibfnamefont {G.}~\bibnamefont
  {Chiribella}}\ and\ \bibinfo {author} {\bibfnamefont {Y.}~\bibnamefont
  {Yang}},\ }\bibfield  {title} {\bibinfo {title} {Optimal quantum operations
  at zero energy cost},\ }\href@noop {} {\bibfield  {journal} {\bibinfo
  {journal} {Physical Review A}\ }\textbf {\bibinfo {volume} {96}},\ \bibinfo
  {pages} {022327} (\bibinfo {year} {2017})}\BibitemShut {NoStop}%
\bibitem [{\citenamefont {Sparaciari}\ \emph {et~al.}(2017)\citenamefont
  {Sparaciari}, \citenamefont {Oppenheim},\ and\ \citenamefont
  {Fritz}}]{sparaciari2017resource}%
  \BibitemOpen
  \bibfield  {author} {\bibinfo {author} {\bibfnamefont {C.}~\bibnamefont
  {Sparaciari}}, \bibinfo {author} {\bibfnamefont {J.}~\bibnamefont
  {Oppenheim}},\ and\ \bibinfo {author} {\bibfnamefont {T.}~\bibnamefont
  {Fritz}},\ }\bibfield  {title} {\bibinfo {title} {Resource theory for work
  and heat},\ }\href@noop {} {\bibfield  {journal} {\bibinfo  {journal}
  {Physical Review A}\ }\textbf {\bibinfo {volume} {96}},\ \bibinfo {pages}
  {052112} (\bibinfo {year} {2017})}\BibitemShut {NoStop}%
\bibitem [{\citenamefont {Baumgratz}\ \emph {et~al.}(2014)\citenamefont
  {Baumgratz}, \citenamefont {Cramer},\ and\ \citenamefont
  {Plenio}}]{baumgratz2014quantifying}%
  \BibitemOpen
  \bibfield  {author} {\bibinfo {author} {\bibfnamefont {T.}~\bibnamefont
  {Baumgratz}}, \bibinfo {author} {\bibfnamefont {M.}~\bibnamefont {Cramer}},\
  and\ \bibinfo {author} {\bibfnamefont {M.}~\bibnamefont {Plenio}},\
  }\bibfield  {title} {\bibinfo {title} {Quantifying coherence},\ }\href@noop
  {} {\bibfield  {journal} {\bibinfo  {journal} {Physical Review Letters}\
  }\textbf {\bibinfo {volume} {113}},\ \bibinfo {pages} {140401} (\bibinfo
  {year} {2014})}\BibitemShut {NoStop}%
\bibitem [{\citenamefont {Winter}\ and\ \citenamefont
  {Yang}(2016)}]{winter2016operational}%
  \BibitemOpen
  \bibfield  {author} {\bibinfo {author} {\bibfnamefont {A.}~\bibnamefont
  {Winter}}\ and\ \bibinfo {author} {\bibfnamefont {D.}~\bibnamefont {Yang}},\
  }\bibfield  {title} {\bibinfo {title} {Operational resource theory of
  coherence},\ }\href@noop {} {\bibfield  {journal} {\bibinfo  {journal}
  {Physical Review Letters}\ }\textbf {\bibinfo {volume} {116}},\ \bibinfo
  {pages} {120404} (\bibinfo {year} {2016})}\BibitemShut {NoStop}%
\bibitem [{\citenamefont {Marvian}\ and\ \citenamefont
  {Spekkens}(2016)}]{marvian2016quantify}%
  \BibitemOpen
  \bibfield  {author} {\bibinfo {author} {\bibfnamefont {I.}~\bibnamefont
  {Marvian}}\ and\ \bibinfo {author} {\bibfnamefont {R.~W.}\ \bibnamefont
  {Spekkens}},\ }\bibfield  {title} {\bibinfo {title} {How to quantify
  coherence: Distinguishing speakable and unspeakable notions},\ }\href@noop {}
  {\bibfield  {journal} {\bibinfo  {journal} {Physical Review A}\ }\textbf
  {\bibinfo {volume} {94}},\ \bibinfo {pages} {052324} (\bibinfo {year}
  {2016})}\BibitemShut {NoStop}%
\bibitem [{\citenamefont {Streltsov}\ \emph {et~al.}(2017)\citenamefont
  {Streltsov}, \citenamefont {Adesso},\ and\ \citenamefont
  {Plenio}}]{streltsov2017colloquium}%
  \BibitemOpen
  \bibfield  {author} {\bibinfo {author} {\bibfnamefont {A.}~\bibnamefont
  {Streltsov}}, \bibinfo {author} {\bibfnamefont {G.}~\bibnamefont {Adesso}},\
  and\ \bibinfo {author} {\bibfnamefont {M.~B.}\ \bibnamefont {Plenio}},\
  }\bibfield  {title} {\bibinfo {title} {Colloquium: Quantum coherence as a
  resource},\ }\href@noop {} {\bibfield  {journal} {\bibinfo  {journal}
  {Reviews of Modern Physics}\ }\textbf {\bibinfo {volume} {89}},\ \bibinfo
  {pages} {041003} (\bibinfo {year} {2017})}\BibitemShut {NoStop}%
\bibitem [{\citenamefont {Bu\ifmmode~\check{z}\else \v{z}\fi{}ek}\ \emph
  {et~al.}(1999)\citenamefont {Bu\ifmmode~\check{z}\else \v{z}\fi{}ek},
  \citenamefont {Derka},\ and\ \citenamefont {Massar}}]{buzek1999optimal}%
  \BibitemOpen
  \bibfield  {author} {\bibinfo {author} {\bibfnamefont {V.}~\bibnamefont
  {Bu\ifmmode~\check{z}\else \v{z}\fi{}ek}}, \bibinfo {author} {\bibfnamefont
  {R.}~\bibnamefont {Derka}},\ and\ \bibinfo {author} {\bibfnamefont
  {S.}~\bibnamefont {Massar}},\ }\bibfield  {title} {\bibinfo {title} {Optimal
  quantum clocks},\ }\href {https://doi.org/10.1103/PhysRevLett.82.2207}
  {\bibfield  {journal} {\bibinfo  {journal} {Physical Review Letters}\
  }\textbf {\bibinfo {volume} {82}},\ \bibinfo {pages} {2207} (\bibinfo {year}
  {1999})}\BibitemShut {NoStop}%
\bibitem [{\citenamefont {Skrzypczyk}\ \emph {et~al.}(2013)\citenamefont
  {Skrzypczyk}, \citenamefont {Short},\ and\ \citenamefont
  {Popescu}}]{skrzypczyk2013extracting}%
  \BibitemOpen
  \bibfield  {author} {\bibinfo {author} {\bibfnamefont {P.}~\bibnamefont
  {Skrzypczyk}}, \bibinfo {author} {\bibfnamefont {A.~J.}\ \bibnamefont
  {Short}},\ and\ \bibinfo {author} {\bibfnamefont {S.}~\bibnamefont
  {Popescu}},\ }\bibfield  {title} {\bibinfo {title} {Extracting work from
  quantum systems},\ }\href@noop {} {\bibfield  {journal} {\bibinfo  {journal}
  {arXiv preprint arXiv:1302.2811}\ } (\bibinfo {year} {2013})}\BibitemShut
  {NoStop}%
\bibitem [{\citenamefont {{\AA}berg}(2014)}]{aaberg2014catalytic}%
  \BibitemOpen
  \bibfield  {author} {\bibinfo {author} {\bibfnamefont {J.}~\bibnamefont
  {{\AA}berg}},\ }\bibfield  {title} {\bibinfo {title} {Catalytic coherence},\
  }\href@noop {} {\bibfield  {journal} {\bibinfo  {journal} {Physical Review
  Letters}\ }\textbf {\bibinfo {volume} {113}},\ \bibinfo {pages} {150402}
  (\bibinfo {year} {2014})}\BibitemShut {NoStop}%
\bibitem [{\citenamefont {Navascu{\'e}s}\ and\ \citenamefont
  {Popescu}(2014)}]{navascues2014energy}%
  \BibitemOpen
  \bibfield  {author} {\bibinfo {author} {\bibfnamefont {M.}~\bibnamefont
  {Navascu{\'e}s}}\ and\ \bibinfo {author} {\bibfnamefont {S.}~\bibnamefont
  {Popescu}},\ }\bibfield  {title} {\bibinfo {title} {How energy conservation
  limits our measurements},\ }\href@noop {} {\bibfield  {journal} {\bibinfo
  {journal} {Physical Review Letters}\ }\textbf {\bibinfo {volume} {112}},\
  \bibinfo {pages} {140502} (\bibinfo {year} {2014})}\BibitemShut {NoStop}%
\bibitem [{\citenamefont {Koch}\ \emph {et~al.}(2007)\citenamefont {Koch},
  \citenamefont {Terri}, \citenamefont {Gambetta}, \citenamefont {Houck},
  \citenamefont {Schuster}, \citenamefont {Majer}, \citenamefont {Blais},
  \citenamefont {Devoret}, \citenamefont {Girvin},\ and\ \citenamefont
  {Schoelkopf}}]{koch2007charge}%
  \BibitemOpen
  \bibfield  {author} {\bibinfo {author} {\bibfnamefont {J.}~\bibnamefont
  {Koch}}, \bibinfo {author} {\bibfnamefont {M.~Y.}\ \bibnamefont {Terri}},
  \bibinfo {author} {\bibfnamefont {J.}~\bibnamefont {Gambetta}}, \bibinfo
  {author} {\bibfnamefont {A.~A.}\ \bibnamefont {Houck}}, \bibinfo {author}
  {\bibfnamefont {D.}~\bibnamefont {Schuster}}, \bibinfo {author}
  {\bibfnamefont {J.}~\bibnamefont {Majer}}, \bibinfo {author} {\bibfnamefont
  {A.}~\bibnamefont {Blais}}, \bibinfo {author} {\bibfnamefont {M.~H.}\
  \bibnamefont {Devoret}}, \bibinfo {author} {\bibfnamefont {S.~M.}\
  \bibnamefont {Girvin}},\ and\ \bibinfo {author} {\bibfnamefont {R.~J.}\
  \bibnamefont {Schoelkopf}},\ }\bibfield  {title} {\bibinfo {title}
  {Charge-insensitive qubit design derived from the cooper pair box},\
  }\href@noop {} {\bibfield  {journal} {\bibinfo  {journal} {Physical Review
  A}\ }\textbf {\bibinfo {volume} {76}},\ \bibinfo {pages} {042319} (\bibinfo
  {year} {2007})}\BibitemShut {NoStop}%
\bibitem [{\citenamefont {Kjaergaard}\ \emph {et~al.}(2020)\citenamefont
  {Kjaergaard}, \citenamefont {Schwartz}, \citenamefont {Braum{\"u}ller},
  \citenamefont {Krantz}, \citenamefont {Wang}, \citenamefont {Gustavsson},\
  and\ \citenamefont {Oliver}}]{kjaergaard2020superconducting}%
  \BibitemOpen
  \bibfield  {author} {\bibinfo {author} {\bibfnamefont {M.}~\bibnamefont
  {Kjaergaard}}, \bibinfo {author} {\bibfnamefont {M.~E.}\ \bibnamefont
  {Schwartz}}, \bibinfo {author} {\bibfnamefont {J.}~\bibnamefont
  {Braum{\"u}ller}}, \bibinfo {author} {\bibfnamefont {P.}~\bibnamefont
  {Krantz}}, \bibinfo {author} {\bibfnamefont {J.~I.-J.}\ \bibnamefont {Wang}},
  \bibinfo {author} {\bibfnamefont {S.}~\bibnamefont {Gustavsson}},\ and\
  \bibinfo {author} {\bibfnamefont {W.~D.}\ \bibnamefont {Oliver}},\ }\bibfield
   {title} {\bibinfo {title} {Superconducting qubits: Current state of play},\
  }\href@noop {} {\bibfield  {journal} {\bibinfo  {journal} {Annual Review of
  Condensed Matter Physics}\ }\textbf {\bibinfo {volume} {11}},\ \bibinfo
  {pages} {369} (\bibinfo {year} {2020})}\BibitemShut {NoStop}%
\bibitem [{\citenamefont {Liu}\ and\ \citenamefont
  {Winter}(2019)}]{liu2019resource}%
  \BibitemOpen
  \bibfield  {author} {\bibinfo {author} {\bibfnamefont {Z.-W.}\ \bibnamefont
  {Liu}}\ and\ \bibinfo {author} {\bibfnamefont {A.}~\bibnamefont {Winter}},\
  }\bibfield  {title} {\bibinfo {title} {Resource theories of quantum channels
  and the universal role of resource erasure},\ }\href@noop {} {\bibfield
  {journal} {\bibinfo  {journal} {arXiv:1904.04201}\ } (\bibinfo {year}
  {2019})}\BibitemShut {NoStop}%
\bibitem [{\citenamefont {Liu}\ and\ \citenamefont
  {Yuan}(2019)}]{liu2019operational}%
  \BibitemOpen
  \bibfield  {author} {\bibinfo {author} {\bibfnamefont {Y.}~\bibnamefont
  {Liu}}\ and\ \bibinfo {author} {\bibfnamefont {X.}~\bibnamefont {Yuan}},\
  }\bibfield  {title} {\bibinfo {title} {Operational resource theory of quantum
  channels},\ }\href@noop {} {\bibfield  {journal} {\bibinfo  {journal}
  {arXiv:1904.02680}\ } (\bibinfo {year} {2019})}\BibitemShut {NoStop}%
\bibitem [{\citenamefont {Paulsen}(2002)}]{paulsen2002completely}%
  \BibitemOpen
  \bibfield  {author} {\bibinfo {author} {\bibfnamefont {V.}~\bibnamefont
  {Paulsen}},\ }\href@noop {} {\emph {\bibinfo {title} {Completely bounded maps
  and operator algebras}}},\ Vol.~\bibinfo {volume} {78}\ (\bibinfo
  {publisher} {Cambridge University Press},\ \bibinfo {year}
  {2002})\BibitemShut {NoStop}%
\bibitem [{\citenamefont {Uhlmann}(1976)}]{uhlmann1976transition}%
  \BibitemOpen
  \bibfield  {author} {\bibinfo {author} {\bibfnamefont {A.}~\bibnamefont
  {Uhlmann}},\ }\bibfield  {title} {\bibinfo {title} {The ``transition
  probability” in the state space of a*-algebra},\ }\href@noop {} {\bibfield
  {journal} {\bibinfo  {journal} {Reports on Mathematical Physics}\ }\textbf
  {\bibinfo {volume} {9}},\ \bibinfo {pages} {273} (\bibinfo {year}
  {1976})}\BibitemShut {NoStop}%
\bibitem [{\citenamefont {Fuchs}\ and\ \citenamefont {Van
  De~Graaf}(1999)}]{fuchs1999cryptographic}%
  \BibitemOpen
  \bibfield  {author} {\bibinfo {author} {\bibfnamefont {C.~A.}\ \bibnamefont
  {Fuchs}}\ and\ \bibinfo {author} {\bibfnamefont {J.}~\bibnamefont {Van
  De~Graaf}},\ }\bibfield  {title} {\bibinfo {title} {Cryptographic
  distinguishability measures for quantum-mechanical states},\ }\href@noop {}
  {\bibfield  {journal} {\bibinfo  {journal} {IEEE Transactions on Information
  Theory}\ }\textbf {\bibinfo {volume} {45}},\ \bibinfo {pages} {1216}
  (\bibinfo {year} {1999})}\BibitemShut {NoStop}%
\bibitem [{\citenamefont {Yadin}\ \emph {et~al.}(2016)\citenamefont {Yadin},
  \citenamefont {Ma}, \citenamefont {Girolami}, \citenamefont {Gu},\ and\
  \citenamefont {Vedral}}]{yadin2016quantum}%
  \BibitemOpen
  \bibfield  {author} {\bibinfo {author} {\bibfnamefont {B.}~\bibnamefont
  {Yadin}}, \bibinfo {author} {\bibfnamefont {J.}~\bibnamefont {Ma}}, \bibinfo
  {author} {\bibfnamefont {D.}~\bibnamefont {Girolami}}, \bibinfo {author}
  {\bibfnamefont {M.}~\bibnamefont {Gu}},\ and\ \bibinfo {author}
  {\bibfnamefont {V.}~\bibnamefont {Vedral}},\ }\bibfield  {title} {\bibinfo
  {title} {Quantum processes which do not use coherence},\ }\href@noop {}
  {\bibfield  {journal} {\bibinfo  {journal} {Physical Review X}\ }\textbf
  {\bibinfo {volume} {6}},\ \bibinfo {pages} {041028} (\bibinfo {year}
  {2016})}\BibitemShut {NoStop}%
\bibitem [{\citenamefont {Chitambar}\ and\ \citenamefont
  {Gour}(2016{\natexlab{a}})}]{chitambar2016critical}%
  \BibitemOpen
  \bibfield  {author} {\bibinfo {author} {\bibfnamefont {E.}~\bibnamefont
  {Chitambar}}\ and\ \bibinfo {author} {\bibfnamefont {G.}~\bibnamefont
  {Gour}},\ }\bibfield  {title} {\bibinfo {title} {Critical examination of
  incoherent operations and a physically consistent resource theory of quantum
  coherence},\ }\href@noop {} {\bibfield  {journal} {\bibinfo  {journal}
  {Physical Review Letters}\ }\textbf {\bibinfo {volume} {117}},\ \bibinfo
  {pages} {030401} (\bibinfo {year} {2016}{\natexlab{a}})}\BibitemShut
  {NoStop}%
\bibitem [{\citenamefont {Chitambar}\ and\ \citenamefont
  {Gour}(2016{\natexlab{b}})}]{chitambar2016comparison}%
  \BibitemOpen
  \bibfield  {author} {\bibinfo {author} {\bibfnamefont {E.}~\bibnamefont
  {Chitambar}}\ and\ \bibinfo {author} {\bibfnamefont {G.}~\bibnamefont
  {Gour}},\ }\bibfield  {title} {\bibinfo {title} {Comparison of incoherent
  operations and measures of coherence},\ }\href@noop {} {\bibfield  {journal}
  {\bibinfo  {journal} {Physical Review A}\ }\textbf {\bibinfo {volume} {94}},\
  \bibinfo {pages} {052336} (\bibinfo {year} {2016}{\natexlab{b}})}\BibitemShut
  {NoStop}%
\bibitem [{\citenamefont {Aberg}(2006)}]{aberg2006quantifying}%
  \BibitemOpen
  \bibfield  {author} {\bibinfo {author} {\bibfnamefont {J.}~\bibnamefont
  {Aberg}},\ }\bibfield  {title} {\bibinfo {title} {Quantifying
  superposition},\ }\href@noop {} {\bibfield  {journal} {\bibinfo  {journal}
  {arXiv preprint quant-ph/0612146}\ } (\bibinfo {year} {2006})}\BibitemShut
  {NoStop}%
\bibitem [{\citenamefont {D{\'\i}az}\ \emph {et~al.}(2018)\citenamefont
  {D{\'\i}az}, \citenamefont {Fang}, \citenamefont {Wang}, \citenamefont
  {Rosati}, \citenamefont {Skotiniotis}, \citenamefont {Calsamiglia},\ and\
  \citenamefont {Winter}}]{diaz2018using}%
  \BibitemOpen
  \bibfield  {author} {\bibinfo {author} {\bibfnamefont {M.~G.}\ \bibnamefont
  {D{\'\i}az}}, \bibinfo {author} {\bibfnamefont {K.}~\bibnamefont {Fang}},
  \bibinfo {author} {\bibfnamefont {X.}~\bibnamefont {Wang}}, \bibinfo {author}
  {\bibfnamefont {M.}~\bibnamefont {Rosati}}, \bibinfo {author} {\bibfnamefont
  {M.}~\bibnamefont {Skotiniotis}}, \bibinfo {author} {\bibfnamefont
  {J.}~\bibnamefont {Calsamiglia}},\ and\ \bibinfo {author} {\bibfnamefont
  {A.}~\bibnamefont {Winter}},\ }\bibfield  {title} {\bibinfo {title} {Using
  and reusing coherence to realize quantum processes},\ }\href@noop {}
  {\bibfield  {journal} {\bibinfo  {journal} {Quantum}\ }\textbf {\bibinfo
  {volume} {2}},\ \bibinfo {pages} {100} (\bibinfo {year} {2018})}\BibitemShut
  {NoStop}%
\bibitem [{\citenamefont {Fannes}(1973)}]{fannes1973continuity}%
  \BibitemOpen
  \bibfield  {author} {\bibinfo {author} {\bibfnamefont {M.}~\bibnamefont
  {Fannes}},\ }\bibfield  {title} {\bibinfo {title} {A continuity property of
  the entropy density for spin lattice systems},\ }\href@noop {} {\bibfield
  {journal} {\bibinfo  {journal} {Communications in Mathematical Physics}\
  }\textbf {\bibinfo {volume} {31}},\ \bibinfo {pages} {291} (\bibinfo {year}
  {1973})}\BibitemShut {NoStop}%
\bibitem [{\citenamefont {Audenaert}(2007)}]{audenaert2007sharp}%
  \BibitemOpen
  \bibfield  {author} {\bibinfo {author} {\bibfnamefont {K.~M.}\ \bibnamefont
  {Audenaert}},\ }\bibfield  {title} {\bibinfo {title} {A sharp continuity
  estimate for the von neumann entropy},\ }\href@noop {} {\bibfield  {journal}
  {\bibinfo  {journal} {Journal of Physics A: Mathematical and Theoretical}\
  }\textbf {\bibinfo {volume} {40}},\ \bibinfo {pages} {8127} (\bibinfo {year}
  {2007})}\BibitemShut {NoStop}%
\bibitem [{\citenamefont {Kretschmann}\ \emph
  {et~al.}(2008{\natexlab{b}})\citenamefont {Kretschmann}, \citenamefont
  {Kribs},\ and\ \citenamefont {Spekkens}}]{kretschmann2008complementarity}%
  \BibitemOpen
  \bibfield  {author} {\bibinfo {author} {\bibfnamefont {D.}~\bibnamefont
  {Kretschmann}}, \bibinfo {author} {\bibfnamefont {D.~W.}\ \bibnamefont
  {Kribs}},\ and\ \bibinfo {author} {\bibfnamefont {R.~W.}\ \bibnamefont
  {Spekkens}},\ }\bibfield  {title} {\bibinfo {title} {Complementarity of
  private and correctable subsystems in quantum cryptography and error
  correction},\ }\href@noop {} {\bibfield  {journal} {\bibinfo  {journal}
  {Physical Review A}\ }\textbf {\bibinfo {volume} {78}},\ \bibinfo {pages}
  {032330} (\bibinfo {year} {2008}{\natexlab{b}})}\BibitemShut {NoStop}%
\end{thebibliography}%

%

\end{document}